\documentclass[journal]{IEEEtran}

\pdfoutput=1
\usepackage{amsthm}
\usepackage{amssymb}
\usepackage{amsfonts}
\usepackage{amsfonts,subfigure,multicol,color,verbatim,graphicx,cite,epsfig,amssymb,amsmath,cases,bm,xcolor,multirow,array}
\usepackage{bm,algorithm,algorithmicx,algpseudocode}
\usepackage{multirow}
\usepackage{multicol}
\usepackage{supertabular}
\usepackage{array}
\usepackage{colortbl}
\usepackage{longtable}
\usepackage{subfigure}
\usepackage{dsfont}
\usepackage{url}

\usepackage{epstopdf,booktabs}
\usepackage[justification=centering]{caption}
\allowdisplaybreaks[4]

\usepackage{soul} 
\usepackage{color, xcolor} 
\soulregister{\cite}7 
\soulregister{\citep}7 
\soulregister{\citet}7 
\soulregister{\ref}7 
\soulregister{\pageref}7 
\soulregister{\eqref}7  



\renewcommand\appendix{\par
    \setcounter{section}{0}
    \setcounter{subsection}{0}
    \gdef\thesection{Appendix \Alph{section}}}

\begin{document}
\title{ Performance analysis of satellite-terrestrial integrated radio access networks based on stochastic geometry
}

\author{Yaohua~Sun,~ Ruiwen~Li
	\thanks{Yaohua Sun (sunyaohua@bupt.edu.cn), and Ruiwen~Li (lrw2023110189@bupt.edu.cn) are with the State Key Laboratory of Networking and Switching Technology, Beijing University of Posts and Telecommunications, Beijing 100876, China. }}

\maketitle

\begin{abstract}

To enhance coverage and improve service continuity, satellite-terrestrial integrated radio access network (STIRAN) has been seen as an essential trend in the development of 6G. However, there is still lack of theoretical analysis on its coverage performance. To fill this gap, we first establish system model to characterize a typical scenario where low-earth-orbit(LEO) satellites and terrestrial base stations are both deployed. Then, stochastic geometry is utilized to analyze the downlink coverage probability under the setting of shared frequency and distinct frequencies.
Specifically, we derive mathematical expressions for the distances distribution from the serving station to the typical user and the associated probability based on the maximum bias power selection strategy (Max-BPR). Taking into account real-world satellite antenna beamforming patterns in two system scenarios, we derive the downlink coverage probabilities in terms of parameters such as base station density and orbital inclination. Finally, the correctness of the theoretical derivations is verified through experimental simulations, and the influence of network design parameters on the downlink coverage probability is analyzed.


\end{abstract}

\section{introduction}
With the commercialization and technology promotion of
the fifth-generation (5G) cellular networks, global research and exploration for the next generation mobile communication technology (6G) have been initiated.
Compared to mobile communication technologies up to 5G, 6G is expected to achieve worldwide wireless connectivity without any gaps. It aims to provide heterogeneous services and seamless network coverage for everyone and everything\cite{34}. However, terrestrial cellular communication networks unable to realize the aforementioned requirements independently\cite{33}, \cite{40}. Satellite communication networks provide a direct solution to the coverage issue with the wide coverage ability. As a promising paradigm for the development of 6G, STIRANs can greatly extend network coverage and reduce the dependence on terrestrial infrastructure\cite{31}. The International Telecommunication Union (ITU), the 3rd Generation Partnership Project (3GPP), and the IMT-2030 (6G) Promotion Group have conducted research on performance evaluation, key technologies, and network architecture for STIRANs.

As for network performance analysis, authors in \cite{35} compare Telesat, OneWeb, and Starlink satellite communication systems in terms of constellation orbits, communication frequency bands, beams, and link budgets. In \cite{36}, authors utilize system-level simulation to analyze the communication capacity and regional capacity density of a single LEO satellite in the 5G NR system. However, the aforementioned studies primarily utilize numerical simulation methods. When there is a large number of network nodes,  challenges arise in terms of low simulation efficiency. Additionally, it is not very tractable to characterize how design parameters affect network performance. As a widely used performance analysis tool in wireless network research, stochastic geometry can effectively address this challenge.
\subsection{Related Work}
Stochastic geometry has been utilized in both terrestrial cellular communication networks and satellite communication networks. When it comes to terrestrial cellular networks, the Poisson point process (PPP) is the key to modeling the locations of terrestrial base stations (BSs) using stochastic geometry tools. This allows for the approximation of the probability density function (PDF) of the serving distance, leading to analytical expressions for coverage probability, data rate, ergodic capacity, etc.For the downlink communication scenario, the single-layer network coverage probability is analyzed in \cite{1}. In \cite{3}, authors study downlink coverage probability and area spectral efficiency under the line-of-sight (LOS)/non-line-of-sight (NLOS) path-loss model. Extensions to multi-tier case , authors in \cite{4} study the downlink coverage probability and average data rate in a multi-tier network which consists of K independent tiers of PPP distributed BSs. In \cite{5}, the average rate of downlink heterogeneous cellular networks is derived, where a maximum biased-received-power based tier association strategy is presented.For the uplink case, in \cite{2}, the authors investigate  uplink transmission with fractional channel inversion power control. In \cite{6}, authors investigate the coverage probability of multi-layer cellular networks. Further, the more realistic scenarios is studied In \cite{7}, the impact of local and ambient environmental shading on the uplink coverage probability is analyzed. The studies mentioned above demonstrate the effectiveness and analytical tractability of the stochastic geometric approach in studying terrestrial cellular network performance. This approach provides valuable insights into the impact of network parameters and can guide BSs deployment.

In recent times, with the rapid development of LEO communication, there has been a growing interest in analyzing satellite network performance using stochastic geometry tools. One popular approach is to model the spatial distribution of satellite locations using a homogeneous binomial point process (BPP) on the sphere. Based on this, in \cite{8}, authors derive the analytical expressions for coverage probability and data rate, where the effective number of satellites is introduced for matching the uneven satellite intensity across latitudes in the actual constellation. In \cite{9}, authors further investigate the mathematical expression of the effective number of satellites. In \cite{10}, the distance distribution between the user and the nearest satellite is characterized, where a multi-layer LEO satellite constellation is considered. According to \cite{10}, authors in \cite{11} jointly study the coverage probability of both the user-gateway link and the gateway-satellite link to derive end-to-end coverage probability, which provide a more comprehensive understanding of satellite network performance. In \cite{12}, authors study the user downlink outage probability and throughput, whereas directional beamforming with fixed satellite beam antennas is employed without the inter-satellite interference. In \cite{13}, authors investigate the impact of atmospheric attenuation on downlink coverage probability. And to assess the accuracy of the modeling method, authors in \cite{14} employed the Wasserstein Distance-inspired approach to analyze the similarity between the Fibonacci grid and the orbital model.
It's worth noting that modeling satellite distribution with PPP is also a widely used technique.
In \cite{15}, the downlink coverage probability based on the LOS/NLOS transmission channel model is analyzed by using the contact angle distribution. In \cite{18} and \cite{19}, the coverage probability is investigated, where the optimal average number of satellites that maximize coverage performance is obtained. However, the beamwidth control issue was not considered in these studies. In \cite{16} and \cite{17}, directional beamforming is adopted to derive expressions for downlink coverage probability, where the satellite beam width is optimized to maximize network throughput. What's more, other important properties in satellite networks to be considered. In \cite{20}, authors investigate the moments and distribution of the conditional coverage probability to analyze the user fairness. Nevertheless, the impact of satellite distribution at different latitudes is considered in existing work.To address the impact of satellite distribution at different latitudes, several studies have employed the Nonhomogeneous Poisson point process (NPPP) to model satellite positions \cite{21} and \cite{22}. These studies have also analyzed inter-satellites interference, providing a more accurate representation of the spatial distribution of satellites.

Compared with downlink communication, there are few studies on uplink communication. As for the uplink communication, in \cite{23}, authors focus on the coverage probability in frequency multiplexing scenarios, where a constant satellite antenna gain is considered. However, many existing works assume a constant satellite gain and do not account for the actual beamforming pattern. This assumption can lead to discrepancies between theoretical analysis and real outcomes. In \cite{24}, authors utilized actual antenna orientation maps to assess the outage probability and ergodic capacity. Despite these advancements, there is still a gap in understanding the coverage performance for specific orbit geometries. To address this issue, authors in \cite{25} developed a model using a homogeneous PPP for fixed circular orbits and derived the downlink coverage probability for the typical user.

Further, some recent researchers have begun to analyze the performance of satellite-aerial networks \cite{26}-\cite{28}, STIRANs\cite{29},\cite{30}, etc.In \cite{29}, authors derive the expressions for the downlink outage probability and regional spectrum in the coexistence scenario of Geosynchronous Earth Orbit (GEO) satellites and terrestrial cellular communication networks.In \cite{30}, the uplink coverage probability is obtained in the STIRANs scenario with constant antenna gains.

While these prior works provide valuable insights into the coverage performance of densely deployed LEO satellite networks, they have not fully considered the impact of satellite orbit parameters. Furthermore, the majority of these studies' models assume a fixed satellite gain, which may not reflect the actual beamforming patterns. Further research is required to construct an appropriate model for STIRANs and analyze its downlink coverage performance.
\subsection{Contributions}
\par
Further, some recent researchers have begun to analyze the performance of satellite-aerial networks. To address the limitations in prior studies and achieve a more comprehensive understanding of coverage performance, we focus on the impact of satellite orbit parameters. Specifically, we analyze STIRANs scenario and present an analytical approach to model the downlink performance. In our model, satellite locations are modeled as a one-dimensional PPP on orbit, BSs are treated as a homogeneous PPP where the distance from a typical user to BSs is greater than $R_0$. The expression for the downlink coverage probability is derived in terms of parameters such as base station density and orbital inclination. The main contributions of this paper are summarized as follows:

\begin{itemize}
\item The system model accounts for actual satellite antenna beamforming patterns, the one-dimension PPP distribution of satellites, and the PPP distribution of BSs. Users can freely select access networks. The analysis considers two scenarios: one where the typical user communicates with satellite and terrestrial networks that do not share the frequency spectrum, and another where they communicate on the same frequency spectrum.

\item We derive the satellite's visible probability and the conditional distance distribution from the typical user to the serving station. Based on this distance distribution, we determine the probabilities of a typical user connecting to a satellite and a terrestrial base station. Additionally, we provide analytical expressions for the downlink coverage probabilities of STIRANs communication under two system scenarios, respectively.

\item
    Simulations validate the accuracy of the derived coverage probability expressions. When comparing STIRANs with satellite and terrestrial networks, the results indicate that STIRANs coverage probability outperforms both under specific scenarios. We also investigate the impact of SINR threshold, satellite and BS density, orbit inclination, and $R_0$ on STIRANs coverage probability.
\end{itemize}
\par The remainder of the paper is organized as follows. Section \ref{System_Model} describes the proposed network and the propagation model. In Section \ref{satellite_visible_probability}, satellite visible probability is derived. Section \ref{different frequencies} computes the analytical expression for the downlink coverage probabilities under the setting of distinct frequencies. In section \ref{same_frequency}, the analytical expressions for the downlink coverage probabilities under the setting of same frequency is derived. The simulation results and detailed discussions are offered in Section \ref{results}. Finally, we conclude this paper in Section \ref{conclusion}.
\section{System Model} \label{System_Model}
In this section, we explain the proposed network model and propagation model of STIRANs.
\subsection{Network Model}\label{Network_Model}
\subsubsection{Typical user}
We assume the typical user is located at $u =\left(0,0,R_{\mathrm{E}}\right)$ within a Cartesian coordinate system and is equipped with an omnidirectional antenna. According to \cite{37}, by altering the orbit geometries in relation to the specific user location, our analysis framework is suitable for various locations. In light of the user's location, our analysis framework is universal. Assuming that the typical user can receive signals simultaneously from both the satellite and the terrestrial BS. In STIRANs with different frequency downlink communication, the typical user adopts the Max-BRP association strategy to select the access network.
\subsubsection{Satellite network}
\begin{figure}[h]
	\centering
    \subfigure[]{
	\includegraphics[width=0.35\textwidth]{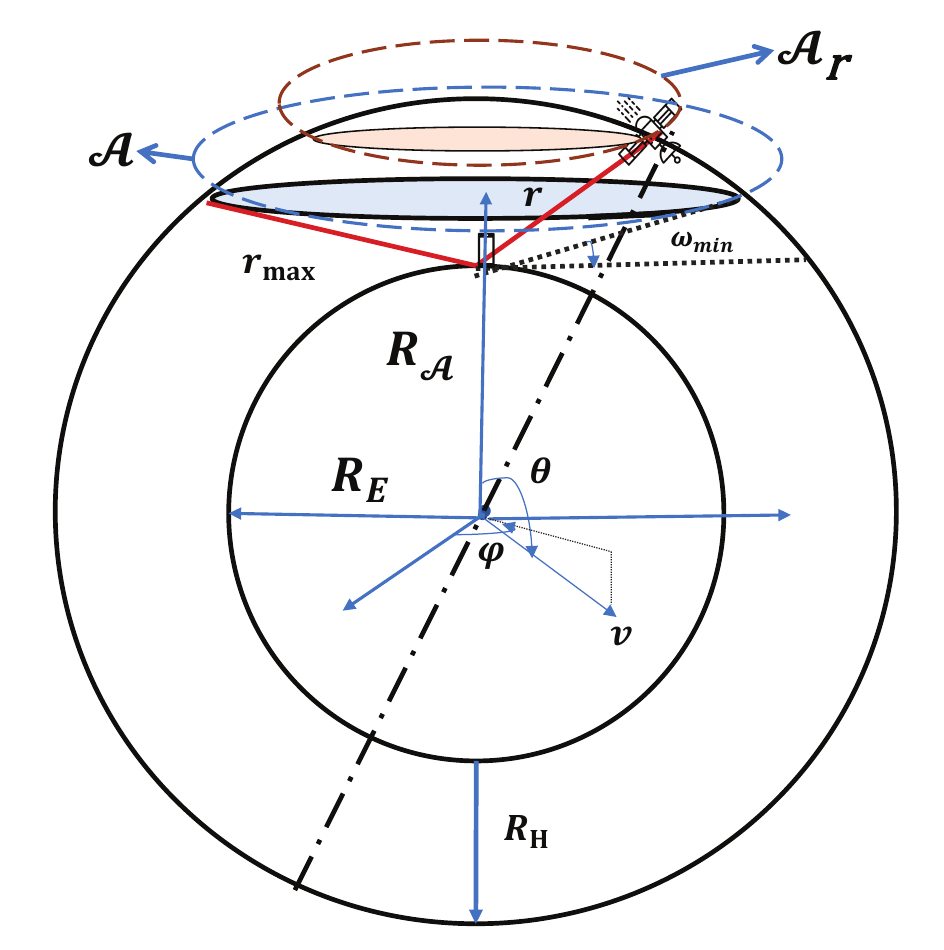}
    }
    \subfigure[]{
	\includegraphics[width=0.35\textwidth]{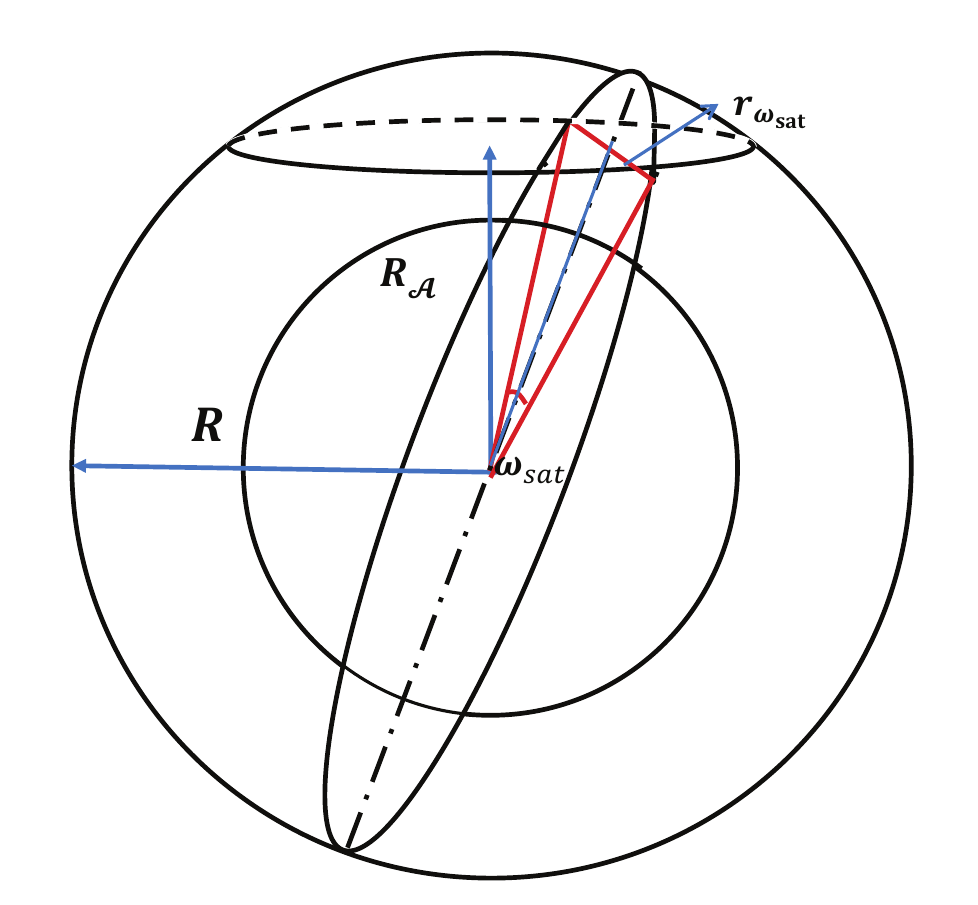}
    }
	\caption{LEO network model}
	\label{fig:LEO network_model}
\end{figure}

As shown in Fig. \ref{fig:LEO network_model}(a), to facilitate analysis, we consider a single LEO orbit satellite communication system with the altitude $R_{\mathrm{H}}$.The satellite orbit is regarded as a concentric circle with the center of the earth as the origin and a radius of $R$, where $R = R_{\mathrm{E}}+R_{\mathrm{H}}$ with $R_{\mathrm{E}}$ represents the Earth's radius.
 The orbit geometry is completely determined by a normal unit-vector $\bar{v} = (1,\theta,\phi)$ with respect to the plane containing the orbit. Denoted the orbit as $\mathcal{O} $, $\theta \in [0,\pi]$ is polar angle, and $\phi \in[0,2\pi]$ is azimuth angle. The satellites are uniformly distributed on an orbit according to a homogeneous PPP with density $\lambda_{\mathrm{sat}}$ and denote the set of their locations as $\bm{\varPhi_{\mathrm{sat}}} = \{x_{\mathrm{sat},1},x_{\mathrm{sat},2},\ldots,x_{\mathrm{sat},M_{\mathrm{sat}}} \}$. The number of satellites on orbit $M_{\mathrm{sat}}$ follows a Poisson distributed with mean $2\pi R$.
\begin{equation}
\mathbb{P}\left( M_{\mathrm{sat}}=\mathrm{m} \right) =\exp \left( -2\pi R\lambda _{\mathrm{sat}} \right) \frac{\left( 2\pi \lambda _{\mathrm{sat}} \right) ^m}{m!}.
\end{equation}
In Fig. \ref{fig:LEO network_model}(a), the minimum elevation angle is $\omega_{\min}$ . The typical user could be served by satellites that are located above the $\omega_{\min}$ and the visible sphere observed by the typical user is represented by $\mathcal{A}$.
The region $\mathcal{A}$ in the Cartesian coordinate is given by
\begin{equation}\label{eq::A}
\mathcal{A} \,\,=\,\,\left\{ \left( x,y,z \right) \in \mathbb{R} ^3:\left\{ x^2+y^2+z^2=R^2 \right\} \cap \left\{ z>R_{\mathcal{A}} \right\} \right\}.
\end{equation}
Where $R_{\mathcal{A}}$ is the distance from the center of the Earth to the base of region $\mathcal{A} $, which is given by
\begin{equation}\label{eq::R::A}
R_{\mathcal{A}}  =r_{\textrm{max}}\sin{\omega_{min}}+R_{\mathrm{E}}
\end{equation}
with $r_{\textrm{max}}$ and $r_{\textrm{min}}$ being the maximum distance between typical user and $\mathcal{A}$, it is given by
\begin{align}
r_{\textrm{max}}&=-R_E\sin\left(\omega_{\min}\right)\nonumber\\
&+\sqrt{R_{\mathrm{E}}^2\sin^2 \left( \omega_{\min}\right)+2R_{\mathrm{E}}R_{\mathrm{H}}+R^2_H}\nonumber
\end{align}
and $r_{\textrm{min}}$ being the minimum distance between typical user and $\mathcal{A}$, it is given by
\begin{align}
r_{\textrm{min}}&=\sqrt{R^2-2R_{\mathrm{E}}Rsin\left(\theta\right)+R_{\mathrm{E}}^2}.
\end{align}
\subsubsection{terrestrial network}
\begin{figure}[h]
	\centering
	\includegraphics[width=0.53\textwidth]{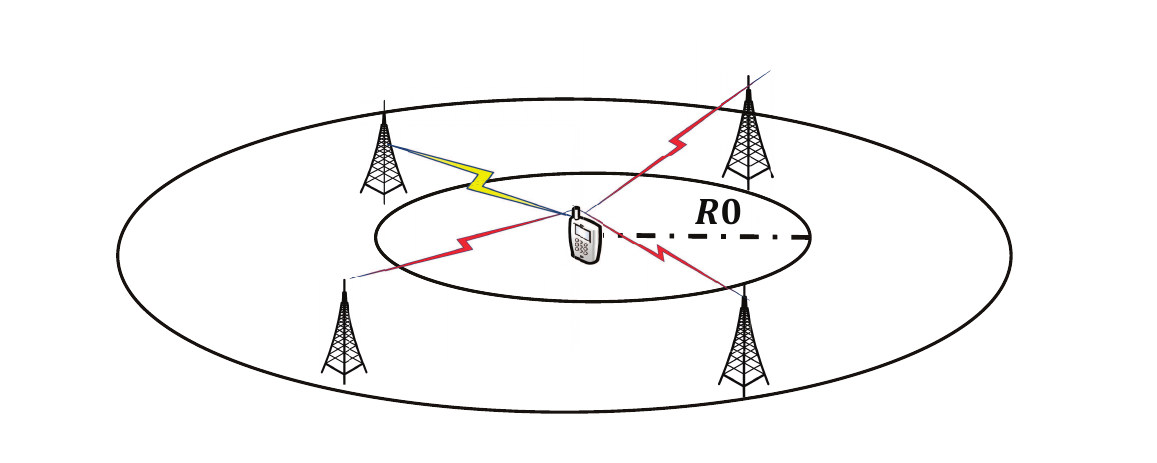}
	\caption{Ground network model}
	\label{fig:Ground network_model}
\end{figure}

For the terrestrial network, we assume that terrestrial BSs are independent of underlying satellites placement. Terrestrial BSs are distributed following a Homogeneous Poisson Point Process (HPPP) with a density $\lambda_{\mathrm{bs}}$. We denote the set of terrestrial BSs locations as $\bm{\varPhi_{\mathrm{bs}}} = \{x_{\mathrm{bs},1},x_{\mathrm{bs},2},\ldots,x_{\mathrm{bs},M_{\mathrm{bs}}} \}$. As shown in Fig.\ref{fig:Ground network_model}, we assume the existence of a coverage hole in terrestrial networks. Within a circle centered on the typical user (depicted as a dot) with a radius $R_0$, there are no BSs located.
\subsection{Propagation Model} \label{Propagation_model}
\subsubsection{communication channel model}
In the realm of wireless communication channels, the wireless channel is composed of two main components: path loss and small-scale fading. Path loss refers to the attenuation or reduction in signal strength as the wireless signal propagates through the wireless channel. Small-scale fading involves rapid signal variations caused by multipath effects, resulting in constructive and destructive interference. In our system, the path loss from the ith satellite to the typical user can be represented as
\begin{align}
r_{\mathrm{sat},i}^{-\alpha_{\mathrm{sat}}} = ||x_{\mathrm{sat},i}-u||^{-\alpha_{\mathrm{sat}}},||x_{\mathrm{sat},i}-u||, x_{\mathrm{sat},i} \in \bm{\varPhi_{\mathrm{sat}}},
\end{align}
where $r_{\mathrm{sat},i}$ is the distance between the user and the i-{th} satellite and $\alpha_{\mathrm{sat}}$ is the path loss exponent for satellite to user communication link. The Shadowed-Rician(SR) fading is used to model the small-scale fading between the terminal and satellites. Denoted the SR fading gain from the $i$-th satellite to the user as $h_{\mathrm{sat},i} \sim \mathrm{SR}(b,m,\varOmega) $. $b$ is the average power of the scattered component, $m$ is the Nakagami fading coefficient $m$ and $\varOmega$ stands for the average power of line-of sight component.Therefore the PDF of $h_{\mathrm{sat},i}$ is given by
 \begin{align}
 f_{h_{\mathrm{sat},i}}\left( x \right)= &\left( \frac{2bm}{2bm+\varOmega} \right) ^m\frac{1}{2b}exp\left( -\frac{x}{2b} \right) \nonumber\\
   & \times F_1\left( m,1,\frac{\varOmega x}{2b\left( 2mb+\varOmega \right)} \right).
 \end{align}
Where $F_1\left(\right)$ is the hypergeometric function. However, the PDF of SR fading is so complex to calculate. According to \cite{38}, the PDF of $h_{\mathrm{sat},i}$ can be tightly approximated by a gamma distribution characterized by parameter $\xi =\frac{m\left( 2b+\Omega \right) ^2}{4mb^2+4mb\Omega +\Omega ^2}$ and $\beta =\frac{4mb^2+4mb\Omega +\Omega ^2}{m\left( 2b+\Omega \right)}$. The PDF of $h_{\mathrm{sat},i}$ is expressed as
\begin{align}
 f_{h_{\mathrm{sat},i}}\left( x \right)= \frac{1}{\beta^\xi \Gamma\left(\xi\right) }x^{\left(\xi-1\right)}\exp\left(-\frac{x}{\beta}\right).
\end{align}
Where $\Gamma(\xi)= \int_{0}^{\inf}{t^\xi\exp{\left(-t\right)}dt}$ is gamma function. The complementary cumulative distribution function (CCDF) of $h_{\mathrm{sat},i}$ can be represented as
\begin{align}
\mathbb{P}\left( h_{\mathrm{sat}}>x \right)&= \int_{x}^{\inf}{\frac{1}{\beta^\xi \Gamma\left(\xi\right) }t^{\left(\xi-1\right)}\exp\left(-\frac{t}{\beta}\right)dt}\nonumber\\
&\mathop {=}^{(v=\frac{t}{\beta})} \frac{1}{\beta^\xi \Gamma\left(\xi\right)}\int_{\frac{x}{\beta}}^{\inf}{\left(v\beta\right)^{\left(\xi-1\right)}\exp\left(-v\right)dv}\nonumber\\
&\approx 1-\left(1-\exp{\left(-\frac{A}{\beta x}\right)}\right)^{\xi}.
\end{align}
Where $A=\xi (\xi !)^{-\frac{1}{\xi!}}$ and  the equality holds when $\xi = 1$. According to the binomial theorem, $\mathbb{P}\left( h_{\mathrm{sat}}>x \right)$ can be represented as
\begin{align}\label{eq::SR}
\mathbb{P}\left( h_{\mathrm{sat}}>x\right)&= \sum_{q=0}^{\xi}{\left( \begin{array}{c}
	\xi\\
	q\\
\end{array} \right)\left( -1\right)^{q+1} \exp{\left(-\frac{A}{\beta x}\right)}}.
\end{align}
\par For the terrestrial communication network, the path loss from the $i$-{th} terrestrial BS to the typical user can be represented as
\begin{align}
r_{\mathrm{bs},i}^{-\alpha_{\mathrm{bs}}} = ||x_{\mathrm{bs},i}-u||^{-\alpha_{\mathrm{bs}}}, x_{\mathrm{bs},i} \in \bm{\varPhi_{\mathrm{bs}}},
\end{align}
where $r_{\mathrm{bs},i}$ is the distance between user and the $i$-th terrestrial BS and $\alpha_{\mathrm{bs}}$ is the path loss exponent for terrestrial BS to user communication link. The Rayleigh fading with mean 1 is used to model the small-scale fading between the user and BSs. Denoted the Rayleigh fading gain from $i$-th BS to the user as $h_{\mathrm{bs},i} \sim \exp(1)$.
\subsubsection{Beamforming gain}
We model the beamforming gain $G(\psi)$ from satellite to user, which is related to the off-axis angle $\psi$. The $\psi$ is given by
\begin{equation}
\psi \left( r \right) =\mathrm{a}rc\cos \left( \frac{R^2+r^2-R_{\mathrm{E}}^{2}}{2Rr} \right).
\end{equation}
Where $r$ is the distance from satellite to user. According to the ITU-R S.1528 LEO \cite{39} reference radiation pattern, the $G(\psi)$ is given by
\begin{equation}
G\left( \psi \right)=\left\{
\begin{array}{lcl}
G_{\mathrm{max}}       &      & {\psi <\psi _b}\\
G_{\mathrm{max}}–3(\frac{\psi}{\psi _b})^2     &      & {\psi _b<\psi <Y}\\
G_{\mathrm{max}}+L_S–25\log\mathrm{(}\frac{\psi}{Y})     &      & {Y<\psi <Z}\\
L_F       &      & {Z<\psi}
\end{array}. \right.
\end{equation}
Where $G_{\mathrm{max}}$ is maximum gain in the main lobe, $L_{\mathrm{S}}$ is main beam and near-in-side-lobe mask cross point(dB) below ideal patterns, $L_{\mathrm{F}}$ is far-out-side-lobe level, $\psi_b$ is one half the 3 dB beamwidth in the plane of interest at the largest off-axis angle, $Y=\psi _b\left( -\frac{L_{\mathrm{S}}}{3} \right) ^{\frac{1}{2}}$ and $Z=Y\times 10^{0.04}\left(G_{\mathrm{max}}+L_{\mathrm{S}}+L_{\mathrm{F}} \right) $.
\section{Satellite Visible Probability} \label{satellite_visible_probability}
To calculate the network coverage probability, it's essential to determine the distribution of the distance to the nearest satellite using the satellite visible probability. In this section, we derive the satellite visible probability.
\par In order to determine the satellite's visibility probability, we first calculate the length of its visible orbit trajectory, which denoted as $L\left( R, \theta, R_{\mathcal{A}} \right)$ in terms of the satellite's orbital parameters. When $\cos\left| \frac{\pi}{2}-\theta \right|>\frac{R_{\mathcal{A}}}{R}$, the satellite orbit does not pass through the $\mathcal{A}$, so $L\left( R,\theta ,R_{\mathcal{A}} \right)=0$; As for $\cos \left| \frac{\pi}{2}-\theta \right|<\frac{R_{\mathcal{A}}}{R}$, $L\left( R,\theta ,R_{\mathcal{A}} \right)$ can be determined by the multiplication of R and $\omega_{\mathrm{sat}}$. $\omega_{\mathrm{sat}}$ can be calculated using the isosceles triangle in Fig.\ref{fig:LEO network_model}(b) and the law of cosines. Which is given by
\begin{align}
\omega_{\mathrm{sat}}
&=\arccos(\frac{R^2+R^2-r^2_{\omega_{sat}}}{2R^2}),
\end{align}
where $r_{\omega_{\mathrm{sat}}} = 2\sqrt{R^2-{R_{\mathcal{A}}}^2(\frac{1}{\sin\theta})^2}$. Meanwhile, the expression of the length of visible orbit trajectory is calculated by
\begin{align}\label{eq::length}
&L\left(R,\theta ,R_{\mathcal{A}} \right)\nonumber\\
&=R\omega_{\mathrm{sat}}\nonumber\\
&=\left\{ \begin{array}{cl}
R\arccos\left(\eta \left( R,\theta ,R_{\mathrm{A}}\right) \right)
 & \cos \left| \frac{\pi}{2}-\theta \right|\le \frac{R_{\mathcal{A}}}{R}\\
	0&\cos \left| \frac{\pi}{2}-\theta \right|>\frac{R_{\mathcal{A}}}{R}\\
\end{array}, \right .
\end{align}
where
\begin{equation}\label{eq::eta::PPP}
\eta \left( R,\theta ,R_{\mathrm{A}}\right)  =\frac{{R_{\mathcal{A}}}^2}{2R^2\sin \!\:{\theta }^2}-1 .
\end{equation}
Denote the probability of the existence of at least one satellite on the visible trajectory is the satellite visible probability $P_{\mathrm{vis}}$. Based on the probability calculation formula of Poisson distribution, the satellite visible probability $P_{\mathrm{vis}}$ can be expressed as
\begin{align}\label{eq::P:vis}
\mathrm{P}_{\mathrm{vis}}&=\mathbb{P}\left( M\left( L\left( R,\theta ,R_{\mathcal{A}} \right) \right) >0 \right)\nonumber\\
&=1-\exp({-\lambda L\left( R,\theta ,R_{\mathcal{A}} \right)}).
\end{align}
Where $M\left( L\left( R,\theta ,R_{\mathcal{A}} \right) \right)$ is the number of satellites on the visible orbit.
\section{The Downlink Coverage Probability In STIRANs with Different Frequencies} \label{different frequencies}
In this section, we compute the nearest satellite/terrestrial BS distance distribution. Based on this, we provide the downlink coverage probability in STIRANs with different frequencies.
\subsection{Distance Distribution} \label{distance distribution}
The statistical properties of distance are the foundation of providing the expressions for coverage probability. In this subsection, we derive the conditional complementary cumulative distribution function (CCDF) and probability density function (PDF) of the distance from the typical user to the serving station.
\subsubsection{The conditional distribution of the nearest satellite distance}
 We define the serving satellite as the nearest satellite to the typical user. Conditioned on that there exists at least one satellite on the visible orbit, we firstly calculate CCDF of the distance from typical user to the nearest satellite.
 Let $r_{\mathrm{sat}}$ is the distance from the nearest satellite to typical user.
 When $r_{\mathrm{sat}} = r$, the spherical cap region denoted as $\mathcal{A}_{r}$ by taking $r$ as the maximum distance is formed as shown in Fig.\ref{fig:LEO network_model}(1). The distance from typical user to the satellites in $\mathcal{A}_{r}$ is smaller than $r$ and $\mathcal{A}_{r}\cap \mathcal{O}$ means the intersection between $\mathcal{A}_{r}$ and the satellite's orbit.
 We calculate the distance from the Earth center to the base of $\mathcal{A}_r$ first, which is $\frac{\left( R^2+R_{\mathrm{E}}^{2}-r^2 \right)}{2R_{\mathrm{E}}}$. Then, similar to (\ref{eq::length}) and (\ref{eq::eta::PPP}), the $L\left( \mathcal{A} _r\cap \mathcal{O} \right)$ is given by
\begin{equation}\label{eq:L:r:sat}
L\left( \mathcal{A} _r\cap \mathcal{O}  \right) =R\arccos\!\:\left( \eta \left( R,\theta ,\frac{\left( R^2+R_{\mathrm{E}}^{2}-r^2 \right)}{2R_{\mathrm{E}}} \right) \right).
\end{equation}
 The probability of $r_{\mathrm{sat}}< r$ is the probability that there is no satellites on the visible orbit in $\mathcal{A}_{r} $, conditioned $ M\left( L\left( R,\theta  ,R_{\mathcal{A}} \right) \right) >0$, is given by
\begin{equation}\label{eq:F:r:sat:proof}
\begin{split}
&F_{r_{\mathrm{sat}}}\left( r \right)\\
&=\mathbb{P}\left( r_{\mathrm{sat}}>r \middle| M\left( L\left( R,\theta ,R_{\mathcal{A}} \right) \right) >0 \right)
\\
&=\frac{\mathbb{P}\left( M\left( L\left( \mathcal{A} _r\cap \mathcal{O} \right) \right) =0,M\left( L\left( R,\theta ,R_{\mathcal{A}} \right) \right) >0 \right)}{\mathbb{P}\left( M\left( L\left( R,\theta ,R_{\mathcal{A}} \right) \right) >0 \right)}
\\
&\mathop {=}^{(a)} \frac{\mathbb{P}\left( M\left( L\left( \mathcal{A} _r\cap \mathcal{O} \right) \right) =0 \right)}{\mathbb{P}\left(M\left( L\left( R,\theta ,R_{\mathcal{A}} \right) \right) >0 \right)}\\
&\quad \times \frac{\mathbb{P}\left( M\left( L\left(\mathcal{A} _r\cap \mathcal{O} \right) /L\left( R,\theta ,R_{\mathcal{A}} \right) \right) >0 \right)}{\mathbb{P}\left(M\left( L\left( R,\theta ,R_{\mathcal{A}} \right) \right) >0 \right)}
\\
&\mathop {=}^{(b)}\frac{\exp{ \left(-\lambda_{\mathrm{sat}} L\left( \mathcal{A} _r\cap \mathcal{O}\right) \right)}}{ 1-\exp{ \left( -\lambda_{\mathrm{sat}} L\left( R,\theta ,R_{\mathcal{A}} \right) \right) }}\\
 & \quad \times \frac{ 1-\exp{\left( \left( -\lambda_{\mathrm{sat}} |L\left( \mathcal{A} _r\cap \mathcal{O} \right) /L\left( R,\theta ,R_{\mathcal{A}} \right) |>0\right) \right)}}{ 1-\exp{ \left( -\lambda_{\mathrm{sat}} L\left( R,\theta ,R_{\mathcal{A}} \right) \right) }}
\\
&\mathop {=}\frac{\exp{-\left(\lambda_{\mathrm{sat}} L\left( \mathcal{A} _r\cap \mathcal{O} \right)\right)} -\exp{(-\lambda_{\mathrm{sat}} L\left( R,\theta ,R_{\mathcal{A}} \right)) }}{  1-\exp{ \left( -\lambda_{\mathrm{sat}} L\left( R,\theta ,R_{\mathcal{A}} \right) \right) } }.
\end{split}
\end{equation}
Where ($a$) comes from the independence of the PPP for non-overlapping areas, ($b$) follows from void probability of PPP. In order to obtain the expression, we need to compute the length of the intersection between $\mathcal{A}_{r}$ and the satellite's orbit $L\left( \mathcal{A} _r\cap \mathcal{O}  \right)$ .

Plugging (\ref{eq:L:r:sat}) into (\ref{eq:F:r:sat:proof}), we obtain the conditional CCDF expression of $r_{\mathrm{sat}}$, it is given by
\begin{align}\label{eq:F:r:sat}
F_{r_{\mathrm{sat}}}\left( r \right)=&
\frac{\exp \left( -\lambda_{\mathrm{sat}} R\arccos\left( \eta \left( R,\theta,\frac{R^2+R_{E}^{2}-{r}^2}{2R_E} \right) \right) \right)}{1-\exp\left( -\lambda_{\mathrm{sat}} L\left( R,\theta,R_{\mathcal{A}} \right) \right)}\nonumber\\
&-\frac{\exp \left( -\lambda_{\mathrm{sat}} L\left( R,\theta,R_{\mathcal{A}} \right) \right)}{1-\exp\left( -\lambda_{\mathrm{sat}} L\left( R,\theta,R_{\mathcal{A}} \right) \right)}.
\end{align}
Finally, the PDF of $r_{\mathrm{sat}}$ conditioned on at least one satellite exists on the visible trajectory can be derived by taking derivative of $1-F_{r_{\mathrm{sat}}}\left( r\right) $,which is given by (\ref{eq:f:r:sat}) for $r_{\mathrm{min}}< r <r_{\mathrm{max}}$, while $f_{r_{\mathrm{sat}}}\left( r \right)  = 0$ for otherwise.

\begin{figure*}[http]

\begin{align}\label{eq:f:r:sat}
f_{r_{\mathrm{sat}}}\left( r \right) =\frac{\exp\left(-\lambda_{\mathrm{sat}} R\arccos \left( \eta \left( R,\theta,\frac{R^2+R^{2}_{E}-r^2}{2R_E} \right) \right) \right)2\pi \lambda_{\mathrm{sat}} \left(R^2+R_{E}^{2}-r^2\right)}{{RR^2_E}\sin^2 {\theta}\left(1-\exp \left( -\lambda_{\mathrm{sat}} L\left( R,\theta,R_{\mathcal{A}} \right)\sqrt{1-\left( \eta \left( R,\theta,\frac{R^2+R_{E}^{2}-r^2}{2R_E} \right) \right) ^2}\right)\right)}
\end{align}
\end{figure*}
\subsubsection{ The distribution of the nearest terrestrial BS distance }
\par As for terrestrial network, we define the nearest terrestrial BS as the serving terrestrial BS. According to the terrestrial network model, the CCDF of distance from the typical user to the nearest terrestrial BS can be given by
\begin{equation}\label{eq:F:r:bs}
\begin{split}
F_{r_{\mathrm{bs}}}\left(r\right)&=\mathbb{P}\left(r_{\mathrm{bs}}>r\right)\\
&=\exp \left( -\pi \lambda _{\mathrm{bs}}\left( r^2-R_{0}^{2} \right) \right),  \left( r > R_0 \right).
\end{split}
\end{equation}
Similar to (\ref{eq:f:r:sat}), the PDF of $r_{bs}$ can be derived by taking derivative of $1-F_{r_{\mathrm{bs}}}$. It is given by
\begin{equation}\label{eq:f:r:bs}
f_{r_{\mathrm{bs}}}(r)=2\pi \lambda _{\mathrm{bs}}(r)\exp\mathrm{(}-\pi \lambda _{\mathrm{bs}}(r^2-{R_0}^2)), \left( r > R_0 \right).
\end{equation}

\subsection{Coverage Probability of Downlink Network }
In this subsection, precise formula is presented for the downlink coverage probabilities in STIRANs with different frequencies. The coverage probability is defined as the probability of $\mathrm{SINR}>\bar{\gamma}$, where $\mathrm{SINR}$ is received signal-to-interference-plus-noise ratio and $\bar{\gamma}$ is the given threshold. The downlink coverage probability of STIRANs with different frequencies is formally defined as
\begin{align}\label{eq:P:cov_diff}
\mathrm{P}^{\mathrm{cov}}_{\mathrm{diff}}
&= 1-(1-\mathrm{P}_{\mathrm{diff,sat}}^{\mathrm{cov}}) \times (1-\mathrm{P}_{\mathrm{diff,bs}}^{\mathrm{cov}}).
\end{align}
If the SINR of either the satellite exceeds a threshold $\bar{\gamma}$ or the SINR of the terrestrial BS surpasses $\bar{\gamma}$ , the transmission will succeed. Where $\mathrm{P}_{\mathrm{diff,sat}}^{\mathrm{cov}}$ and $\mathrm{P}_{\mathrm{diff,bs}}^{\mathrm{cov}}$ are the downlink coverage probability in satellite networks and terrestrial networks, respectively.
\subsubsection{The Downlink Coverage Probability Of The Satellite network}
\par In regards to the satellite network, the coverage probability of the typical user is derived as
\begin{equation} \label{eq:P:diff_sat:cov}
\begin{split}
\mathrm{P}_{\mathrm{sat}}^{\mathrm{cov}} &=\mathrm{P}_{\mathrm{vis}} \times \mathrm{P}_{\mathrm{sat|M>0}}^{\mathrm{cov}}\\
&=\mathrm{P}_{\mathrm{vis}} \times \mathbb{E}\left\{ \mathbb{P}\left(\mathrm{SINR}^{\mathrm{diff}}_{\mathrm{sat}}>\bar{\gamma}|r_{\mathrm{sat}},M>0 \right) \right\}.
\end{split}
\end{equation}
$\mathrm{P}_{\mathrm{vis}}$ is obtained by  (\ref{eq::P:vis}), and $\mathrm{P}_{\mathrm{diff,sat|M>0}}^{\mathrm{cov}}$ is the downlink coverage probability conditioned that at least one satellite exists on the visible trajectory. $\mathrm{SINR}^{\mathrm{diff}}_{\mathrm{sat}}>\bar{\gamma}$ represents
the SINR of the received signal at the satellite exceeds a given threshold $T$.
$S_{\mathrm{sat}}$ is the received power of the serving satellite, which is given by
\begin{equation} \label{eq:S_sat}
S_{\mathrm{sat}} = G\left( r_{\mathrm{sat}} \right) P_{\mathrm{t,sat}}r^{-\alpha _{\mathrm{sat}}}_{\mathrm{sat}}h_{\mathrm{sat}},
\end{equation}
where $P_{\mathrm{t,sat}}$ is the transmit power of satellites. $r_{\mathrm{sat}}$ signifies the distance between the serving satellite and the typical user, $h_{\mathrm{sat}}$ represents the small-scale fading coefficient between the serving satellite  and the typical user, $G\left( r_{\mathrm{sat}} \right)$ is the receive antenna gain between the serving satellite  and the typical user.
SINR at the receiver from the perspective of the satellite in STIRANs with different frequencies, which can be expressed as
\begin{equation} \label{eq:SINR:diff_sat}
\mathrm{SINR}^{\mathrm{diff}}_{\mathrm{sat}}=\frac{S_{\mathrm{sat}}}{ I_{\mathrm{sat}}+\sigma ^2 }=\frac{G\left( r_{\mathrm{sat}} \right) P_{t,\mathrm{sat}}r^{-\alpha _{\mathrm{sat}}}_{\mathrm{sat}}h_{\mathrm{sat}}}{ I_{\mathrm{sat}}+\sigma ^2 },
\end{equation}
where $\sigma ^2 $ denotes the variance of the noise and $I_{\mathrm{sat}}$ is the aggregated interference power from satellites, which is given by
\begin{equation} \label{eq:I:sat}
I_{\mathrm{sat}}=\sum_{i\in \bm{\varPhi _{\mathrm{sat}}}}{G\left( r_{\mathrm{sat},i}\right) P_{t,\mathrm{sat}}r^{-\alpha _{\mathrm{sat}}}_{\mathrm{sat},i}{h_{\mathrm{sat},i}}}.
\end{equation}
where $G\left( r_{\mathrm{sat},i} \right)$ is the  receive antenna gain between the $i$-th interfering satellite and the typical user.
Therefore, the downlink coverage probability conditioned on at least one satellite exists on the visible trajectory orbit can be derived in
\begin{align} \label{eq:P:cov:sat:diff}
&\mathrm{P}_{\mathrm{diff,sat|M>0}}^{\mathrm{cov}}\nonumber\\
&=\mathbb{E}\left\{ \mathbb{P}\left(\mathrm{SINR}^{\mathrm{diff}}_{\mathrm{sat}}>\bar{\gamma}|r_{\mathrm{sat}} \right) \right\}\nonumber\\
&=\int_{r_{\mathrm{min}}}^{r_{\mathrm{max}}}{\mathbb{E}\left(\mathbb{P}\left( h_{\mathrm{sat}}>\frac{\bar{\gamma}\left( I_{\mathrm{sat}}+\sigma ^2 \right) {r}^{\alpha _{\mathrm{sat}}}}{G\left( r\right) P_{t,\mathrm{sat}}} \right) \right) f_{r_{\mathrm{sat}}}(r)dr}\nonumber\\
&\mathop {=}^{(a)}\int_{r_{\mathrm{min}}}^{r_{\mathrm{max}}}{\sum_{q=0}^{\xi}{\left( \begin{array}{c}
	\xi\\
	q\\
\end{array} \right)} e^{-\mathrm{s_{\mathrm{sat}}}\sigma ^2}\mathcal{L} _{I_{\mathrm{sat}}}\left( s_{\mathrm{sat}}\right) (-1)^qf_{r_{\mathrm{sat}}}(r)dr}.
\end{align}
Where $s_{\mathrm{sat}} =\frac{A\bar{\gamma}r^{\alpha_{\mathrm{sat}}}q}{\beta G\left( r\right) P_{t,\mathrm{sat}}}$. According to (\ref{eq::SR}), (a) follows from the binomial theorem and the tight upper bound of a gamma random variable of parameters $\beta$ and $\xi$.
Conditioned that the nearest satellite is located with the distance of $r_{\mathrm{sat}} = r $, We compute the Laplace transform of the aggregated interference power of inter-satellite. Such Laplace transform is computed as
\begin{equation} \label{eq:L:sat:sat}
\begin{split}
&\mathcal{L} _{I_{\mathrm{sat}}}\left( s_{\mathrm{sat}} \right) \\
&=\mathbb{E}_{I_{\mathrm{sat}}}\left\{\exp\left({-s_{\mathrm{sat}} I_{\mathrm{sat}}}\right) \right\}
\\
&=\mathbb{E}_{I_{\mathrm{sat}}}\left\{ \exp\left({-s_{\mathrm{sat}} \sum_{i\in \bm{\varPhi} _{\mathrm{sat}}}{G\left( r_{\mathrm{sat,i}} \right)P_{\mathrm{t,sat}}{r^{-\alpha _{\mathrm{sat}}}_{\mathrm{sat,i}}}h_{\mathrm{sat},i}}}\right) \right\}
\\
&\mathop {=}^{(a)}\exp{\left(\int_r^{r_{\mathrm{max}}}{\left(\mathbb{E}\left(\frac{s_{\mathrm{sat}} P_{\mathrm{t,sat}}h_{\mathrm{sat}}G\left( u \right)}{ u^{\alpha _{\mathrm{sat}}}}\right )-1\right)f\left( u \right) du}\right)}
\\
&\mathop {=}^{(b)}\exp{\left( -\int_r^{r_{\mathrm{max}}}{\left( 1-\left( \frac{1}{1+\frac{s_{\mathrm{sat}} \beta P_{\mathrm{t,sat}}G\left( u \right)}{ u^{\alpha _{\mathrm{sat}}}}} \right) ^{\xi} \right)f\left( u \right) du}\right) }.
\end{split}
\end{equation}
Where (a) follows from the probability generating functional (PGFL) of PPP, (b) comes from the independence of the small-scale fading from the point process and leverages the gamma distribution of channel fading. By change of variables with respect to the distance from the typical user, $f\left(u \right) $ is obtained given by
\begin{equation}\label{eq:f:u}
f\left( u \right) =\frac{2\lambda _{\mathrm{sat}}u\left( R^2+R^{2}_{E}-u^2 \right)}{{RR^2_E}\sin^2 {\theta} \sqrt{1-\left( \eta \left( R,\theta,\frac{R^2+R_{E}^{2}-u^2}{2R_E} \right) \right) ^2} }.
\end{equation}
\subsubsection{The Downlink Coverage Probability of The Terrestrial network}
As for terrestrial networks, the donwlink coverage probability of the typical user is derived as
\begin{equation}\label{eq:P:diff_bs:cov}
\mathrm{P}_{\mathrm{diff,bs}}^{\mathrm{cov}}=\mathbb{E}\left\{ \mathbb{P}\left(\mathrm{SINR}^{\mathrm{diff}}_{\mathrm{bs}}>\bar{\gamma}|r_{\mathrm{bs}} \right) \right\}.
\end{equation}
$\mathrm{SINR}^{\mathrm{diff}}_{\mathrm{bs}}>\bar{\gamma}$ represents
the SINR of the received signal at the terrestrial BS exceeds a given threshold $\bar{\gamma}$. $S_{\mathrm{bs}}$ is the received power of the serving terrestrial BS, which is given by
\begin{equation} \label{eq:S_bs}
S_{\mathrm{bs}} = P_{\mathrm{t,bs}}{r^{-\alpha _{\mathrm{bz}}}_{\mathrm{bs}}}h_{\mathrm{bs}},
\end{equation}
where $P_{\mathrm{t,bs}}$ is the transmit power of terrestrial BSs. $r_{\mathrm{bs}}$ signifies the distance between the serving terrestiral BS and the typical user, $h_{\mathrm{bs}}$ represents the small-scale fading coefficient between the serving terrestrial BS and the typical user.
$\mathrm{SINR}^{\mathrm{diff}}_{\mathrm{bs}}$ represents
SINR at the receiver from the perspective of the terrestrial network in STIRANs with different frequencies, which can be expressed as
\begin{equation} \label{eq:SINR:diff_bs}
\mathrm{SINR}^{\mathrm{diff}}_{\mathrm{bs}}=\frac{S_{\mathrm{bs}}}{ I_{\mathrm{bs}}+\sigma ^2 }=\frac{P_{t,\mathrm{bs}}{r^{-\alpha _{\mathrm{bz}}}_{\mathrm{bs}}}h_{\mathrm{bs}}}{I_{\mathrm{bs}}+\sigma ^2 },
\end{equation}
where  $I_{\mathrm{bs}}$ is the aggregated interference power from interfering terrestrial BSs, which is given by
\begin{equation} \label{eq:I:bs}
I_{\mathrm{bs}}=\sum_{i\in \bm{\varPhi _{\mathrm{bs}}}}{P_{t,\mathrm{bs}}{r^{-\alpha _{\mathrm{bz}}}_{\mathrm{bs},i}}{h_{\mathrm{bs},i}}}.
\end{equation}
Similarly, the downlink coverage probability under $r_{\mathrm{bs}} = r $ is derived as
\begin{equation}\label{eq:P:cov:bs:diff:proof}
\begin{split}
&\mathrm{P}_{\mathrm{diff,bs}}^{\mathrm{cov}}\\
&= \mathbb{E}\left\{ \mathbb{P}\left( \mathrm{SINR}^{\mathrm{diff}}_{\mathrm{bs}}>\bar{\gamma} \right) |r_{\mathrm{bs}} \right\}\\
&= \int_{R_0}^{\inf}{} \mathbb{E}\left\{\mathbb{P}\left( h_{\mathrm{bs}}>\frac{\bar{\gamma}\left( I_{\mathrm{bs}}+\sigma ^2 \right) {r}^{\alpha_{\mathrm{bs}}}}{P_{\mathrm{t,bs}}} \right)\right\} f_{r_{\mathrm{bs}}}(r)dr\\
&\mathop  = \limits^{(a)}\int_{R_0}^{\inf}{ \exp {(-s_{\mathrm{bs}} \sigma ^2)} \mathcal{L} _{I_{\mathrm{bs}}}\left( s_{\mathrm{bs}}  \right)f_{r_{\mathrm{bs}}}(r)dr}.
\end{split}
\end{equation}
 Where $s_{\mathrm{bs}} = \frac{\bar{\gamma}}{r_{\mathrm{bs}}^{-\alpha_{\mathrm{bs}}}P_{t,\mathrm{bs}}}$, (a) using the fact that $h_{\mathrm{bs}} \sim \exp \left(1\right)$. $\mathcal{L} _{I_{\mathrm{bs}}}\left( s_{\mathrm{bs}} \right)$ is the Laplace transform of cumulative interference power $I_{\mathrm{bs}}$. Conditioned that the nearest satellite is located with the distance of $r_{\mathrm{bs}} = r $, it can be expressed as
\begin{equation} \label{eq:L:bs:bs}
\begin{split}
&\mathcal{L} _{I_{\mathrm{bs}}}\left( s_{\mathrm{bs}} \right) \\ &=\mathbb{E}_{I_{\mathrm{bs}}}\left( \exp{\left( -s_{\mathrm{bs}} I_{\mathrm{bs}}\right)}\right)
\\
&=\mathbb{E}_{I_{\mathrm{bs}}}\left( \exp{(-s_{\mathrm{bs}} \sum_{i\in \bm{\varPhi _{\mathrm{bs}}}}{P_{\mathrm{t,bs}}{r^{-\alpha _{\mathrm{bs}}}_{\mathrm{bs},i}}h_{\mathrm{bs},i}})} \right)
\\
&\mathop {=}^{(a)}\exp{\left( -2\pi \lambda _{\mathrm{bs}}\int_r^{\inf}{\left(1-\mathbb{E}_{h_{\mathrm{bs}}}\left(s_{\mathrm{bs}} P_{\mathrm{t,bs}}{u}^{-\alpha _{\mathrm{bs}}}h_{\mathrm{bs}} \right) \right)u du}\right)}
\\
&\mathop {=}^{(b)}\exp{\left(-2\pi \lambda _{\mathrm{bs}}\int_r^{inf}{\left( 1-\left( \frac{1}{1+\frac{s_{\mathrm{bs}} P_{t,\mathrm{bs}}}{u^{\alpha _{\mathrm{bs}}}}} \right)\right)udu}\right)},
\end{split}
\end{equation}
where (a) follows from the PGFL of PPP, (b) comes from the independence of the small-scale fading from the point process and $h_{\mathrm{bs},i} \sim \exp(1)$ .
\section{The Downlink Coverage Probability In STIRANs with the Same Frequency} \label{same_frequency}
In STIRANs with a same frequency scenario, satellite and terrestrial transmissions share the same frequency and interference between satellite and terrestrial is occured. In this section, we first obtain the association probability with the satelites network and the terrestrial network. Based on association probability, we derive the conditional distribution of the nearest satellite/terrestrial BS distance and the downlink coverage probability expressions in STIRANs under a setting of same frequency.
\subsection{Association Probability and Conditional Distance Distribution for Downlink}
We assume that the Max-BRP association strategy is employed, where the receiver selects the desired signal based on the maximum bias received power. In this section, we first derive the association probability based on the Max-BRP association strategy. The conditional distribution of the nearest satellite/terrestrial BS distance is then derived.

Based on the Max-BRP association strategy, we define $\mathrm{sat} = \arg \max\left(S_{\mathrm{bias,n}}\right)$ as the typical user associated with satellite networks and $\mathrm{bs} = \arg \max\left(S_{\mathrm{bs,n}}\right)$ as the typical user associated with terrestrial networks. The biased power received by the typical user from the serving satellite can be represented as
\begin{equation}
S_{\mathrm{bias,sat}} = \rho _{\mathrm{sat}} G\left( r_{\mathrm{sat}} \right) P_{\mathrm{t,sat}}{r^{-\alpha _{\mathrm{sat}}}_{\mathrm{sat}}}h_{\mathrm{sat}}
\end{equation}
and from the serving terrestrial BS can be represented as
\begin{equation}
S_{\mathrm{bias,bs}} = \rho _{\mathrm{bs}}P_{\mathrm{t,bs}}{r}^{-\alpha _{\mathrm{bs}}}_{\mathrm{bs}}h_{\mathrm{bs}},
\end{equation}
where $\rho_{\mathrm{sat}}$ and $\rho_{\mathrm{bs}}$ represent the bias factor of satellite and terrestrial base station, respectively. The association probability between the typical user and satellite is
\begin{small}
\begin{align}\label{eq:P:sat:proof}
&\mathrm{P}_{\mathrm{sat}}=\mathbb{P}(\arg \max\left(S_{\mathrm{bias,n}} = \mathrm{sat}\right))\nonumber\\
&=\mathbb{E}\left\{\mathbb{P}\left(\rho _{\mathrm{sat}}P_{t,\mathrm{sat}}\bar{h}_{\mathrm{sat}}G\left( r_{\mathrm{sat}} \right) r_{\mathrm{sat}}^{-\alpha _{\mathrm{sat}}}>\rho _{\mathrm{bs}}P_{t,\mathrm{bs}}r_{\mathrm{bs}}^{-\alpha _{\mathrm{bs}}}|r_{\mathrm{sat}} \right) \right\}\nonumber\\
&=\mathbb{E}\left\{\mathbb{P}\left(r_{\mathrm{bs}}>\left( \frac{\rho _{\mathrm{bs}}P_{t,\mathrm{bs}}}{\rho _{\mathrm{sat}}P_{t,\mathrm{sat}}\bar{h}_{\mathrm{sat}}G\left( r_{\mathrm{sat}} \right) r_{\mathrm{sat}}^{-\alpha _{\mathrm{sat}}}} \right) ^{\frac{1}{\alpha _{\mathrm{bs}}}}|r_{\mathrm{sat}} \right) \right\}\nonumber\\
&=\int_{r_{\mathrm{min}}}^{r_{\mathrm{max}}}{F_{\mathrm{b}} \times f_{r_{\mathrm{sat}}}\left( r \right) dr}.
\end{align}
\end{small}
The mathematical expectation of SR fading channel gain is $\bar{h}_{\mathrm{sat}}$, The mathematical expectation of rayleigh fading channel gain is 1. $F_{\mathrm{b}} $ is the probability of maximum biased received power of satellite larger than terrestrial base station under the distance from the typical user to and nearest satellite is $r$. According to (\ref{eq:F:r:bs}), the $F_{\mathrm{b}} $ can be expressed as
\begin{small}
\begin{equation} \label{eq:F_down_b}
\begin{split}
&F_{\mathrm{b}}=\mathbb{P}\left(r_{\mathrm{bs}}>\left( \frac{\rho _{\mathrm{bs}}P_{\mathrm{t,bs}}}{\rho _{\mathrm{sat}}P_{\mathrm{t,sat}}\bar{h}_{\mathrm{sat}}G\left( r \right) {r}^{-\alpha _{\mathrm{sat}}}} \right) ^{\frac{1}{\alpha _{\mathrm{bs}}}} \right)\\
&=\exp{ \left(-\pi \lambda _\mathrm{bs}\left(\left(\frac{\rho _{bs}P_{\mathrm{t,bs}}}{\rho _{\mathrm{sat}}P_{\mathrm{t,sat}}\bar{h}_{\mathrm{sat}}G\left(r\right) {r}^{-\alpha _{\mathrm{sat}}}} \right) ^{\frac{2}{\alpha _{\mathrm{bs}}}}-R_{0}^{2} \right)\right)},
\end{split}
\end{equation}
\end{small}
for $\left( \frac{\rho _{bs}P_{t,\mathrm{bs}}}{\rho _{\mathrm{sat}}P_{t,\mathrm{sat}}\bar{h}G\left( r_{\mathrm{sat}} \right) {r_{\mathrm{sat}}}^{-\alpha _{sat}}} \right) ^{\frac{1}{\alpha _{\mathrm{bs}}}}>R_0$,  and $F_{\mathrm{b}}=1$  for otherwise. Substituting (\ref{eq:F_down_b}) into (\ref{eq:P:sat:proof}), the association probability between the typical user and satellite is expressed by (\ref{eq:P:sat}), where $r_{1}$ can be computed by $\rho _{\mathrm{sat}}P_{t,\mathrm{sat}}\bar{h}_{\mathrm{sat}}G\left(r_{1} \right) {r_{1}}^{-\alpha _{\mathrm{sat}}}=\rho _{\mathrm{bs}}P_{t,\mathrm{bs}}{R_0}^{\alpha _{\mathrm{bs}}}$.
\begin{figure*}[tp]
\begin{align}\label{eq:P:sat}
\mathrm{P}_{\mathrm{sat}}&=\mathbb{P}(\arg \max\left(\rho _{\mathrm{n}}S_{\mathrm{bias,n}} = \mathrm{sat}\right))\nonumber\\
&=\begin{cases}
\int_{r_{\mathrm{min}}}^{r_{\mathrm{max}}}{\exp{\left( -\pi \lambda _{\mathrm{bs}}\left(\left(\frac{\rho _{bs}P_{\mathrm{t,bs}}}{\rho _{\mathrm{sat}}P_{\mathrm{t,sat}}\bar{h}_{\mathrm{sat}}G\left( r \right) r^{-\alpha _{\mathrm{sat}}}} \right) ^{\frac{2}{\alpha _{\mathrm{bs}}}}-R_{0}^{2}\right)\right)} f_{r_{\mathrm{sat}}}\left( r \right) dr}&0<r_{1}\leqslant r_{\mathrm{min}}\\
	\int_{r_{\mathrm{min}}}^{r_{1}}{f_{r_{\mathrm{sat}}}\left( r \right) dr}+\int_{r_{\mathrm{1}}}^{r_{\mathrm{max}}}{\exp{\left(-\pi \lambda _{\mathrm{bs}}\left(\left(\frac{\rho _{bs}P_{\mathrm{t,bs}}}{\rho _{\mathrm{sat}}P_{\mathrm{t,sat}}\bar{h}_{\mathrm{sat}}G\left( r \right) {r}^{-\alpha _{\mathrm{sat}}}} \right) ^{\frac{2}{\alpha _{\mathrm{bs}}}}-R_{0}^{2} \right)\right)} f_{r_{\mathrm{sat}}}\left( r \right) dr}
&r_{\mathrm{min}}<r_{1}\leqslant r_{\mathrm{max}}\\
	\int_{r_{\mathrm{min}}}^{r_{\mathrm{max}}}{f_{r_{\mathrm{sat}}}\left( r \right)dr}                                                                   &r_{\max}\leqslant r_{1}\\
\end{cases}
\end{align}
\end{figure*}

Similarly, the association probability between the typical user and terrestrial BS is
\begin{small}
\begin{align}\label{eq:P:bs}
&\mathrm{P}_{\mathrm{bs}}=\mathbb{P}(\arg \max\left(S_{\mathrm{bias,n}} = \mathrm{bs}\right))\nonumber\\
&=\mathbb{E}\left\{\mathbb{P}\left(\rho _{\mathrm{bs}}P_{t,\mathrm{bs}}{r^{-\alpha _{\mathrm{bs}}}_{\mathrm{bs}}}> \rho_{\mathrm{sat}}P_{t,\mathrm{sat}}\bar{h}_{\mathrm{sat}}G\left( r_{\mathrm{sat}} \right) r_{\mathrm{sat}}^{-\alpha _{\mathrm{sat}}}|r_{\mathrm{bs}}\right) \right\}.
\end{align}
\end{small}
Due to the antenna beamforming gain is a piecewise function of $r_{\mathrm{sat}}$, simplifying (\ref{eq:P:bs}) to resemble the form of (\ref{eq:P:sat:proof}) becomes highly complex. Hence, the association probability between the typical user and terrestrial BS is represented as (\ref{eq:P:bs:proof}). $r_{2}$ and $r_{\mathrm{3}}$ are obtained by $\rho _{\mathrm{sat}}P_{t,\mathrm{sat}}\bar{h}_{\mathrm{sat}}G\left( r_{\mathrm{sat}} \right) r_{\mathrm{min}}^{-\alpha _{\mathrm{sat}}}=\rho _{\mathrm{bs}}P_{t,\mathrm{bs}}{r^{-\alpha _{\mathrm{bs}}}_{\mathrm{2}}}$ and $\rho _{\mathrm{sat}}P_{t,\mathrm{sat}}\bar{h}_{\mathrm{sat}}G\left( r_{\mathrm{sat}} \right) r_{\mathrm{max}}^{-\alpha _{\mathrm{sat}}}=\rho _{\mathrm{bs}}P_{t,\mathrm{bs}}r^{-\alpha _{\mathrm{bs}}}_{\mathrm{3}}$ .
\begin{figure*}[http]
\begin{align}\label{eq:P:bs:proof}
\mathrm{P}_{\mathrm{bs}}&=\mathbb{P}(\arg \max\left(S_{\mathrm{bias,n}} = \mathrm{bs}\right))\nonumber\\
&=\begin{cases}
	0&r_{\mathrm{3}}\leqslant R_0\\
	\int_{R_0}^{r_{\mathrm{3}}}{\mathbb{P}\left( \rho _{\mathrm{sat}}P_{\mathrm{t,sat}}\bar{h}_{\mathrm{sat}}G\left( r_{\mathrm{sat}} \right) {r_{\mathrm{sat}}}^{-\alpha _{\mathrm{sat}}}<\rho _{\mathrm{bs}}P_{\mathrm{t,bs}}r^{-\alpha _{\mathrm{bs}}} \right) f_{r_{\mathrm{bs}}}\left( r \right) dr}&0<r_{\mathrm{2}}<R_0\leqslant d_{\mathrm{3}}\\
	\int_{R_0}^{r_{\mathrm{2}}}{f_{\mathrm{bs}}\left( r \right)}dr+\int_{r_{\mathrm{2}}}^{r_{\mathrm{3}}}{\mathbb{P}\left( \rho _{\mathrm{sat}}P_{\mathrm{t,sat}}\bar{h}_{\mathrm{sat}}G\left( r_{\mathrm{sat}} \right) {r_{\mathrm{sat}}^{-\alpha _{\mathrm{sat}}}}<\rho _{\mathrm{bs}}P_{\mathrm{t,bs}}r^{-\alpha _{\mathrm{bs}}} \right) f_{r_{\mathrm{bs}}}\left( r \right) dr} &0<R_0<r_{\mathrm{2}}<r_{\mathrm{3}}\\
\end{cases}
\end{align}
\end{figure*}
\par When the typical user associates with the satellite, the distance between the typical user and the serving satellite is denoted as $d_{\mathrm{sat}}$ and the CCDF of the $d_{\mathrm{sat}}$ can be expressed as
\begin{equation}\label{eq:F:sat:d}
\begin{split}
F_{d_{\mathrm{sat}}}\left( r\right)=\frac{\mathbb{P}\left(r_{\mathrm{sat}}>r|{n = \mathrm{sat}} \right)}{\mathrm{P}_{\mathrm{sat}}},
\end{split}
\end{equation}
for $r_{\mathrm{min}}<r<r_{\mathrm{max}}$. According to the Max-BRP association strategy, $\mathbb{P}\left(r_{\mathrm{sat}}>r|{n = \mathrm{sat}} \right)$  can be further expressed as
\begin{equation}\label{eq:F:sat:d2}
\begin{split}
 \mathbb{P}\left(r_{\mathrm{sat}}>r|{n = \mathrm{sat}} \right) = \int_{r}^{r_{\mathrm{max}}}{F_{\mathrm{b}}f_{r_{\mathrm{sat}}}\left(u\right) du }.
\end{split}
\end{equation}
Considering on (\ref{eq:F_down_b}), (\ref{eq:P:sat}) and (\ref{eq:F:sat:d2}), the CCDF of the $d_{\mathrm{sat}}$ can be obtained. Meanwhile, the PDF of ${d_{\mathrm{sat}}}$ is $f_{d_{\mathrm{sat}}}(r)$ , which can be computed by taking derivative of $1-F_{d_{\mathrm{sat}}}\left( r\right) $. When $r_{\mathrm{min}}<r<r_{\mathrm{max}}$,  $f_{d_{\mathrm{sat}}}(r)$ is given by

\begin{equation} \label{eq:f:sat:d}
\begin{split}
f_{d_{\mathrm{sat}}}(r)&=\frac{d\left[1-F_{d_{\mathrm{sat}}}\left( r\right)\right]}{dr}\\
&= \frac{\exp\left( -\pi \lambda _{\mathrm{bs}}\left( \left( \frac{\rho _{\mathrm{bs}}P_{\mathrm{t,bs}}}{\rho _{\mathrm{sat}}P_{\mathrm{t,sat}}\bar{h}_{\mathrm{sat}}G\left(r\right){r}^{-\alpha _{\mathrm{sat}}}}\right) ^{\frac{2}{\alpha _{\mathrm{bs}}}}-R_{0}^{2} \right)\right)}{\mathrm{P}_{\mathrm{sat}}}\\
&\quad \times \frac{f_{r_{\mathrm{sat}}}\left(r\right)}{\mathrm{P}_{\mathrm{sat}}},
\end{split}
\end{equation}
for $r>r_1$, and $f_{d_{\mathrm{sat}}}\left(r\right) = f_{r_{\mathrm{sat}}}\left(r\right)$ for $r_1>r$. Meanwhile, the distance between the typical user and serving terrestrial BS is $d_{\mathrm{bs}}$ under the typical user associates with the terrestrial BS. When $r_3<R_0$ and $R_0<r$, the CCDF of $d_{\mathrm{bs}}$ is written by
\begin{equation}\label{eq:f:bs:d}
\begin{split}
&F_{d_{\mathrm{bs}}}\left( r\right)= \frac{\mathbb{P}\left(r_{\mathrm{bs}}>r|{n = \mathrm{bs}} \right)}{\mathrm{P}_{\mathrm{bs}}} \\
&= \frac{\int_{r}^{r_{\mathrm{3}}}{\mathbb{P}\left( \frac{\rho _{\mathrm{sat}}P_{\mathrm{t,sat}}\bar{h}_{\mathrm{sat}}G\left( r_{\mathrm{sat}} \right)} {{r_{\mathrm{sat}}}^{\alpha _{\mathrm{sat}}}}<\frac{\rho _{\mathrm{bs}}P_{\mathrm{t,bs}}}{u^{\alpha _{\mathrm{bs}}}} \right) f_{r_{\mathrm{bs}}}\left( u \right) du}}{\mathrm{P}_{\mathrm{bs}}}
\end{split}
\end{equation}
for $d_2<r<d_3$, and
\begin{equation}\label{eq:f:bs:d}
\begin{split}
&F_{d_{\mathrm{bs}}}\left( r\right)= \frac{\mathbb{P}\left(r_{\mathrm{bs}}>r|{n = \mathrm{bs}} \right)}{\mathrm{P}_{\mathrm{bs}}} \\
&= \frac{\int_{r}^{r_{\mathrm{2}}}{f_{\mathrm{bs}}\left( u \right)}du}{\mathrm{P}_{\mathrm{bs}}}\\&+\frac{\int_{r_{\mathrm{2}}}^{r_{\mathrm{3}}}{\mathbb{P}\left( \frac{\rho _{\mathrm{sat}}P_{\mathrm{t,sat}}\bar{h}_{\mathrm{sat}}G\left( r_{\mathrm{sat}} \right)} {{r_{\mathrm{sat}}}^{\alpha _{\mathrm{sat}}}}<\frac{\rho _{\mathrm{bs}}P_{\mathrm{t,bs}}}{u^{\alpha _{\mathrm{bs}}}} \right) f_{r_{\mathrm{bs}}}\left( u \right) du}}{\mathrm{P}_{\mathrm{bs}}}
\end{split}
\end{equation}
for $r<r_2$. And the $f_{d_{\mathrm{bs}}}(r)$ is computed by taking derivative of $1-F_{d_{\mathrm{bs}}}\left( r\right) $. $f_{d_{\mathrm{bs}}}(r)$ is given by
\begin{equation}\label{eq:f:bs:d}
f_{d_{\mathrm{bs}}}(r)= \frac{{\mathbb{P}\left(\frac{\rho _{\mathrm{sat}}P_{\mathrm{t,sat}}\bar{h}_{\mathrm{sat}}G\left( r_{\mathrm{sat}} \right)}{{r_{\mathrm{sat}}}^{\alpha _{\mathrm{sat}}}}<\frac{\rho _{\mathrm{bs}}P_{\mathrm{t,bs}}}{r^{\alpha _{\mathrm{bs}}}} \right) f_{r_{\mathrm{bs}}}\left( r \right)}}{\mathrm{P}_{\mathrm{bs}}}
\end{equation}
for $r_2<r<r_3$ and $f_{d_{\mathrm{bs}}}\left(r\right) = f_{r_{\mathrm{bs}}}\left(r\right)$ for $r<r_2$.
\subsection{Coverage Probability Of Downlink Transmission}
In this subsection, we apply SG to derive the downlink coverage probability of the whole network in STIRANs with same frequency, which is given by

\begin{align}
\mathrm{P}^{\mathrm{cov}}_{\mathrm{same}}
=&\mathrm{P}_{\mathrm{vis}}\times\left( \mathrm{P}_{\mathrm{sat}}\times \mathrm{P}_{\mathrm{same,sat}}^{\mathrm{cov}}+\mathrm{P}_{\mathrm{bs}}\times \mathrm{P}_{\mathrm{same,bs}}^{\mathrm{cov}}\right)\nonumber\\
&+\mathrm{P}_{\mathrm{invis}}\times \mathrm{P}_{\mathrm{same},\mathrm{invis\_bs}}^{\mathrm{cov}}.
\end{align}

Where $\mathrm{P}_{\mathrm{same,sat}}^{\mathrm{cov}}$ and $\mathrm{P}_{\mathrm{same,bs}}^{\mathrm{cov}}$ are the downlink coverage probability under the typical user associate with satellite and terrestrial BS, respectively. $\mathrm{P}_{\mathrm{invis}}$ is the probability of no satellites on the visible orbit, which is given by $\mathrm{P}_{\mathrm{invis}} = 1-\mathrm{P}_{\mathrm{vis}}$. $\mathrm{P}_{\mathrm{same},\mathrm{invis\_bs}}^{\mathrm{cov}}$ is the downlink coverage probability under no satellite on the visible orbit.

\subsubsection{The Downlink Coverage Probability of the Satellite network}
\par When the typical user associate with satellites, its coverage probability is defined as
\begin{equation} \label{eq:P:same_sat:cov}
\mathrm{P}_{\mathrm{same,sat}}^{\mathrm{cov}}=\mathbb{E}\left\{ \mathbb{P}\left(\mathrm{SINR}^{\mathrm{same}}_{\mathrm{sat}}>\bar{\gamma}|r_{\mathrm{sat}} \right) \right\},
\end{equation}
where $\mathrm{SINR}^{\mathrm{same}}_{\mathrm{sat}}$ represents SINR at the typical user from the perspective of the satellite in STIRANs  with the same frequency, which is given by
\begin{equation} \label{eq:SINR:same_sat}
\mathrm{SINR}^{\mathrm{same}}_{\mathrm{sat}}=\frac{G\left( r_{\mathrm{sat}} \right) P_{\mathrm{t,sat}}{r_{\mathrm{sat}}^{-\alpha _{\mathrm{sat}}}}h_{\mathrm{sat}}}{I_{\mathrm{bs}}+ I_{\mathrm{sat}}+\sigma ^2 }.
\end{equation}

$I_{\mathrm{sat}}$ can be expressed by (\ref{eq:I:sat}) and $I_{\mathrm{bs}}$ is given by (\ref{eq:I:bs}).
Therefore the conditional coverage probability is gven by
\begin{align} \label{eq:P:cov:sat:same:proof}
&\mathrm{P}_{\mathrm{same,bs}}^{\mathrm{cov}}\nonumber\\
&=\mathbb{E}\left\{ \mathbb{P}\left(\mathrm{SINR}^{\mathrm{same}}_{\mathrm{sat}}>\bar{\gamma}|r_{\mathrm{sat}} \right) \right\}\nonumber\\
&=\int_{r_{\mathrm{min}}}^{r_{\mathrm{max}}}{}\mathbb{E}\left(\mathbb{P}\left( h_{\mathrm{sat}}>\frac{\bar{\gamma}\left( I_{\mathrm{bs}}+I_{\mathrm{sat}}+\sigma ^2 \right) {r}^{\alpha _{\mathrm{sat}}}}{G\left(r\right) P_{\mathrm{t,sat}}} \right)\right) f_{d_{\mathrm{sat}}}(r)dr\nonumber\\
&\mathop {=}^{(a)}\int_{r_{\mathrm{min}}}^{r_{\mathrm{max}}}{}\sum_{q=0}^{\xi}{\left( \begin{array}{c}
	\xi\\
	q\\
\end{array} \right)} \exp\left({-\mathrm{s_{\mathrm{sat}}}\sigma ^2}\right)\mathcal{L} _{I_{\mathrm{sat}}}\left( s_{\mathrm{sat}}\right)\nonumber\\
& \qquad \qquad\qquad\qquad \times\mathcal{L} _{I_{\mathrm{bs}}}\left( s_{\mathrm{sat}}\right) (-1)^qf_{d_{\mathrm{sat}}}(r)dr.
 \end{align}
Similar to the (\ref{eq:P:cov:sat:diff}), Where $s_{\mathrm{sat}} =\frac{A\bar{\gamma}r^{\alpha_{\mathrm{sat}}}q}{\beta G\left( r\right) P_{\mathrm{t,sat}}}$, (a) is according to (\ref{eq::SR}).And $\mathcal{L} _{I_{\mathrm{sat}}}(s_{\mathrm{sat}})$ and $\mathcal{L} _{I_{\mathrm{bs}}}(s_{\mathrm{sat}})$ are the Laplace transform of cumulative interference power $I_{\mathrm{sat}}$ and $I_{\mathrm{bs}}$.  When $r_{\mathrm{sat}} = r $, the $\mathcal{L} _{I_{\mathrm{sat}}}(s_{\mathrm{sat}})$ is obtained by (\ref{eq:L:sat:sat}) and  $\mathcal{L} _{I_{\mathrm{bs}}}(s_{\mathrm{sat}})$ is expressed by

\begin{equation}\label{eq:L:sat:bs}
\begin{split}
&\mathcal{L} _{I_{\mathrm{bs}}}\left( s_{\mathrm{sat}} \right) \\
&=\mathbb{E}_{I_{\mathrm{bs}}}\left(e^{ -s_{\mathrm{sat}}I_{\mathrm{bs}}}  \right)
\\
&=\mathbb{E}_{I_{\mathrm{bs}}}\left( e^{-s_{\mathrm{sat}}\sum_{i\in \bm{\varPhi _{\mathrm{bs}}}}{P_{\mathrm{t,bs}}{r^{-\alpha _{\mathrm{bs}}}_{\mathrm{bs,i}}}h_{\mathrm{bs},i}} } \right)
\\
&\mathop {=}^{\left(a\right)}\exp{\left(-\int_r^{r_{\mathrm{max}}}{\left( 1-\mathbb{E}_{h_{\mathrm{bs}}}\left( s_{\mathrm{sat}}P_{\mathrm{t,bs}}u^{-\alpha _{\mathrm{sat}}}h_{\mathrm{bs}} \right) \right) \!f\left( u \right) du}\right)}
\\
&\mathop {=}^{\left(b\right)}\exp{\left(-\int_r^{\inf}{\left( 1-\left( \frac{1}{1+\frac{s_{\mathrm{sat}}P_{\mathrm{t,bs}}}{u^{\alpha _{\mathrm{bs}}}}} \right) \right) f\left( u \right) du}\right)}
\\
&\mathop {=}^{\left(c\right)}\begin{cases}
	e^{ -2\pi \lambda _{\mathrm{bs}}\int_{d_1}^{\inf}{\left( 1-\left( \frac{1}{1+\frac{s_{\mathrm{sat}}P_{\mathrm{t,bs}}}{u^{\alpha _{\mathrm{bs}}}}}  \right) \right) udu} }&		d_1 >R_0\\
	e^{ -2\pi \lambda _{\mathrm{bs}}\int_{R_0}^{\inf}{\left( 1-\left( \frac{1}{1+\frac{s_{\mathrm{sat}}P_{\mathrm{t,bs}}}{u^{\alpha _{\mathrm{bs}}}}}  \right) \right) udu}}&		0<d_1 \leqslant R_0\\
\end{cases}.
\end{split}
\end{equation}
Where $d_1 = \left( \frac{\rho _{\mathrm{bs}}P_{\mathrm{t,bs}}}{\rho _{\mathrm{sat}}P_{\mathrm{t,sat}}\bar{h}G\left( r \right)r^{-\alpha _{\mathrm{sat}}}} \right) ^{\frac{1}{\alpha _{\mathrm{bs}}}}$. (a) follows from the PGFL of PPP, (b) comes from the independence of the small-scale fading, and $f\left(u \right) $ is obtained by (\ref{eq:f:u}). (c) stem form the equation bases the different values of $d_1$ to determine the lower limit of integration.
\subsubsection{Downlink Coverage Probability With Terrestrial BS}
When the typical user communicates with terrestrial BS, its coverage probability is defined as
\begin{equation}
\mathrm{P}_{\mathrm{same,bs}}^{\mathrm{cov}} =\mathbb{E}\left\{ \mathbb{P}\left(\mathrm{SINR}^{\mathrm{same}}_{\mathrm{bs}}>\bar{\gamma}|r_{\mathrm{bs}} \right) \right\}.
\end{equation}
$\mathrm{SINR}^{\mathrm{same}}_{\mathrm{bs}}$ represents SINR at the typical user from the perspective of the terrestrial in STIRANs with same frequency, which is given by
\begin{equation} \label{eq:SINR:bs}
\mathrm{SINR}^{\mathrm{same}}_{\mathrm{bs}}=\frac{P_{t,\mathrm{bs}}{r_{\mathrm{bs}}}^{-\alpha _{\mathrm{bs}}}h_{\mathrm{bs}}}{I_{\mathrm{sat}}+I_{\mathrm{bs}}+\sigma ^2 }.
\end{equation}
Similarly, $I_{\mathrm{sat}}$ and $I_{\mathrm{bs}}$ are given by (\ref{eq:I:sat}) and (\ref{eq:I:bs}), respectively. When $r_{\mathrm{bs}} = r $, the conditional coverage probability under the typical user associate with terrestrial BS is derived as
\begin{align} \label{eq:P:cov:bs:same:proof}
&\mathrm{P}_{\mathrm{same,bs}}^{\mathrm{cov}} = \mathbb{E}\left\{ \mathbb{P}\left( \mathrm{SINR}^{\mathrm{same}}_{\mathrm{bs}}>\bar{\gamma} \right) |r_{\mathrm{bs}} \right\}\nonumber\\
&=\int_{R_0}^{\inf}{ \mathbb{E}\left(\mathbb{P}\left( h_{\mathrm{bs}}>\frac{\bar{\gamma}\left( I_{\mathrm{bs}}+I_{\mathrm{sat}}+\sigma ^2 \right) {r}^{\alpha _{\mathrm{bs}}}}{P_{\mathrm{t,bs}}} \right)\right) f_{d_{\mathrm{bs}}}(r)dr}\nonumber\\
&\mathop  = \limits^{(b)}\int_{R_0}^{\inf}{ e^ {-s_{\mathrm{bs}} \sigma ^2} \mathcal{L} _{I_{\mathrm{bs}}}\left( s_{\mathrm{bs}}  \right)
\mathcal{L} _{I_{\mathrm{sat}}}
\left( s_{\mathrm{bs}}  \right)f_{d_{\mathrm{bs}}}(r)dr}.
\end{align}
(a) comes from the independence of the two events, (b) using the fact that $h_{\mathrm{bs}} \sim \exp \left(1\right)$.
  $\mathcal{L} _{I_{\mathrm{bs}}}\left( s_{\mathrm{bs}} \right)$ is obtained by (\ref{eq:L:bs:bs}) and $\mathcal{L} _{I_{\mathrm{sat}}}\left( s_{\mathrm{bs}}\right)$ can be expressed by
\begin{equation}\label{eq:L:bs:sat}
\begin{split}
&\mathcal{L} _{I_{\mathrm{sat}}}\left( s_{\mathrm{bs}} \right)=\mathbb{E}\left( e^{ -s_{\mathrm{bs}}I_{\mathrm{sat}}}\right)
\\
&=\mathbb{E}_{I_{\mathrm{bs}}}\left( e^{-s_{\mathrm{bs}}\sum_{i\in \varPhi _{\mathrm{sat}}}{G\left(r_{\mathrm{sat},i}\right)P_{\mathrm{t,sat}}r_{\mathrm{sat},i}^{-\alpha _{\mathrm{sat}}}h_{\mathrm{sat},i}} } \right)
\\
&\mathop {=}^{\left(b\right)}\exp{\left(-\int_r^{inf}{\left( 1-\left( \frac{1}{1+\frac{s_{\mathrm{bs}}\beta P_{t,\mathrm{sat}}G\left( u \right)}{u^{\alpha _{\mathrm{sat}}}}} \right) ^{\xi}\right) f\left( u \right) du}\right)}
\\
&\!\!\!\!=\begin{cases}
	e^{\int_{d_2}^{r_{\mathrm{max}}}{\left(\!\left( \frac{1}{1+\frac{s_{\mathrm{bs}}\beta P_{t,\mathrm{sat}}G\left( u \right)}{u^{\alpha _{\mathrm{sat}}}}} \right) ^{\xi}\! -1 \!\right) f\left( u \right) du}}  &\!r_{\mathrm{max}}>d_2>r_{\mathrm{min}}\\
	e^{\int_{r_{\mathrm{min}}}^{r_{\mathrm{max}}}{\left( \left( \! \frac{1}{1+\frac{s_{\mathrm{bs}}\beta P_{\mathrm{t,sat}}G\left( u \right)}{u^{\alpha _{\mathrm{sat}}}}} \right) ^{\xi} \! -1\!\right) f\left( u \right) du}}&\!		r_{\mathrm{min}}>d_2
\end{cases}.
\end{split}
\end{equation}
Where $d_2$ can be obtained by $\rho _{\mathrm{sat}}P_{\mathrm{t,sat}}\bar{h}G\left( d_2 \right) {d_2 }^{-\alpha _{\mathrm{sat}}}=\rho _{\mathrm{bs}}P_{\mathrm{t,bs}}{r}^{-\alpha _{\mathrm{bs}}}$

When there is no satellite on the visible orbit, the STIRANs can be seen as terrestrial networks scenario. The downlink coverage probability under the condition that no satellites existing on the visible trajectory $\mathrm{P}_{\mathrm{same},\mathrm{invis\_bs}}^{\mathrm{cov}}$ can be expressed by
\begin{align} \label{eq:P:cov:bs:in_vis_same}
&\mathrm{P}_{\mathrm{same},\mathrm{invis\_bs}}^{\mathrm{cov}}\nonumber\\
&=\mathbb{E}\left\{ \mathbb{P}\left( \mathrm{SINR}^{\mathrm{same}}_{\mathrm{bs}}>\bar{\gamma} \right) |r_{\mathrm{bs}} \right\}\nonumber\\
&= \int_{R_0}^{\inf}{ \mathbb{E}\left\{\mathbb{P}\left( h_{\mathrm{bs}}>\tfrac{\bar{\gamma}\left( I_{\mathrm{bs}}+\sigma ^2 \right) {r}^{\alpha_{\mathrm{bs}}}}{P_{\mathrm{t,bs}}} \right)\right\} f_{r_{\mathrm{bs}}}(r)dr}\nonumber\\
& = \int_{R_0}^{\inf}{ e^ {-s_{\mathrm{bs}} \sigma ^2} \mathcal{L} _{I_{\mathrm{bs}}}\left( s_{\mathrm{bs}}  \right)
\mathcal{L} _{I_{\mathrm{sat}}}
\left( s_{\mathrm{bs}}  \right)f_{d_{\mathrm{bs}}}(r)dr}.
\end{align}
\section{Numerical results and discussion}  \label{results}
In this section, we provide some numerical examples to verify the derived results based on the simulation parameters listed in Table I unless otherwise stated.
\begin{small}
\begin{table}
\centering
\caption{Simulation parameter setting}
\label{tab:1}
\footnotesize
\begin{tabular}{llllll} \toprule
\textbf{Parameter}                    & \textbf{Notation}  & \textbf{Value} \\
\midrule
Earth’s radius            & $R_E$   &6371km\\
The density of LEO satellites in an orbit   &$\lambda_{\mathrm{sat}}$   & $0.001/ \mathrm{km}$ \\
The density of terrestrial BSs     &$\lambda_{\mathrm{bs}}$ & $1/ \mathrm{km^2}$ \\
Minimum elevation angle  &$\omega_{\mathrm{min}}$ & 10  \\
Satellite transmitter power &$P_{\mathrm{sat}}$& 45 dBm(30W)  \\
Terrestrial BS transmitter power  &$P_{\mathrm{bs}}$   &4 dBm(10W)\\
Orbit altitude                            &$R_H$ & 500 km \\
Orbit inclination              &$\theta$ & $\frac{\pi}{2}$  \\
SINR threshold   & $\bar{\gamma}$     & $-10 \sim 20$ dB\\
Channel fading parameter      &$b$                   & 0.063 \\
Channel fading parameter         &$m$                   & 0.739 \\
Channel fading parameter         &$\varOmega$                   & $8.97\times 10^-4$\\
Maximum receive antenna gain   &$G_{\mathrm{max}}$ &35 dBi \\
System frequency bandwidth    &B & 5 MHz  \\
Noise spectral density          &N  & -174 dBm/Hz   \\
Receiver noise figure            &
  & 11 dB\\
One-half the 3 dB beamwidth &$\varPsi _b$ & 1.6°    \\
Main beam and near-in side-lobe mask\\
cross point (dB) below peak gain       &$L_S$    &-6.75  \\
Far-out side-lobe level &$L_F$ &5 \\
\bottomrule
\end{tabular}
\end{table}
\end{small}
\begin{figure} [http]
    \centering
    \subfigure[$R_0=100\mathrm{m}$]{
	\includegraphics[width=0.427\textwidth]{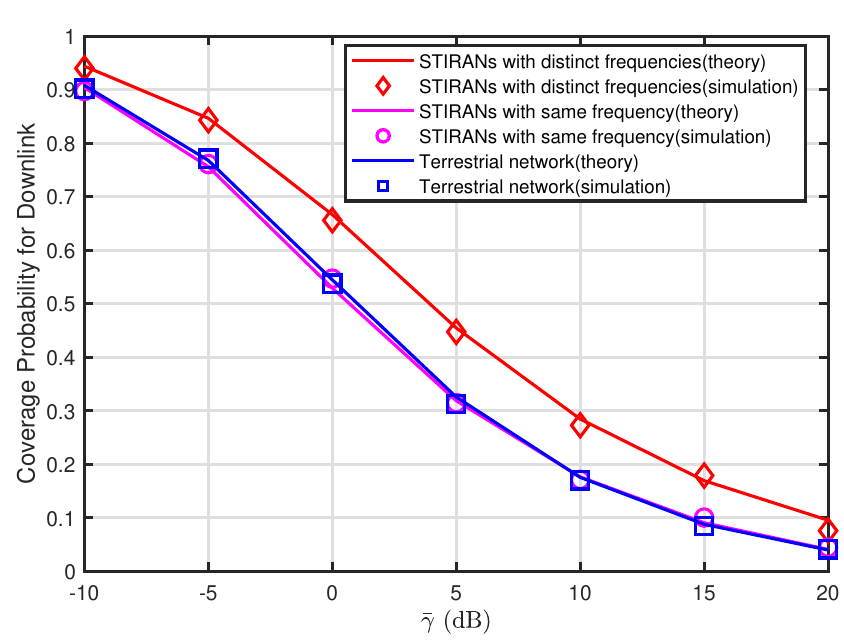}
    }
    \quad
    \subfigure[$R_0 = 700\mathrm{m}$]{
	\includegraphics[width=0.427\textwidth]{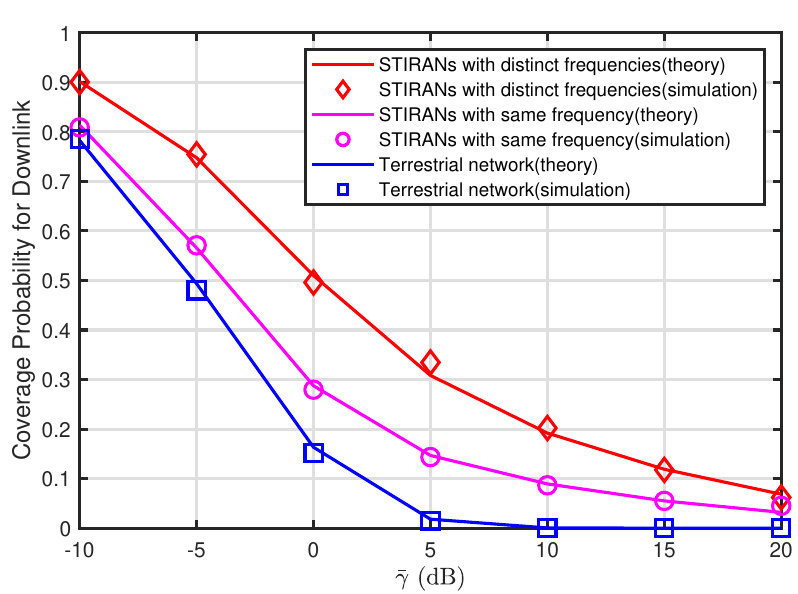}
    }
	\caption{The downlink coverage probability of STIRANs with distinct frequencies, STIRANs with same frequency and terrestrial networks }
	\label{fig:compare}
\end{figure}

In Fig.\ref{fig:compare}, we compare the downlink coverage probability in STIRANs with different frequencies, STIRANs with the same frequency, and terrestrial networks. The accuracy of our analysis is verified through extensive simulation evaluation.
When $R_0=100\mathrm{m}$ and $R_0=700\mathrm{m}$, the downlink coverage probability in STIRANs with different frequencies is significantly higher than in terrestrial networks. The downlink coverage probability of STIRANs with the same frequency is higher than terrestrial networks when $R_0 = 700\mathrm{m}$. Results indicate that the STIRANs effectively solve the coverage problem when there is an existing coverage hole.
\subsection{The downlink probability of the STIRANs with different frequencies}
\begin{figure} [http]
    \centering
    \subfigure[$R_0=100\mathrm{m}$]{
	\includegraphics[width=0.427\textwidth]{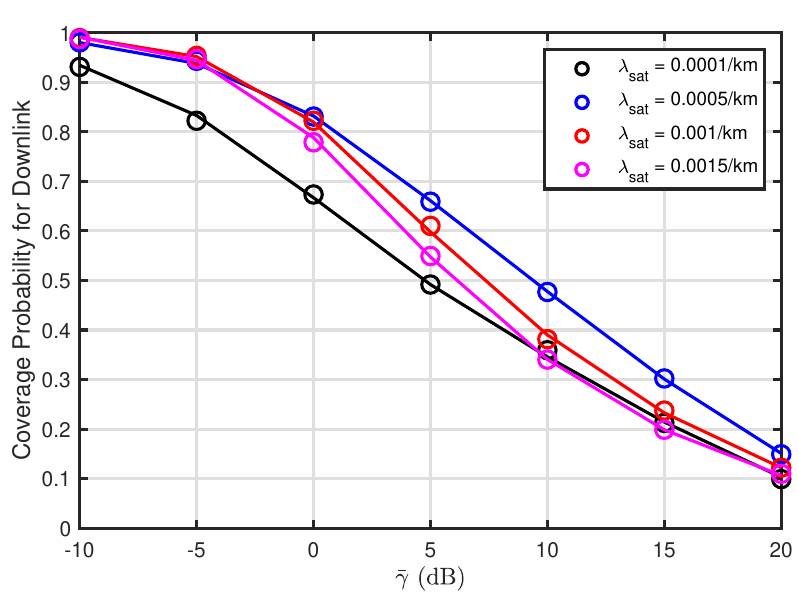}
    }
    \quad
    \subfigure[$R_0 = 700\mathrm{m}$]{
	\includegraphics[width=0.427\textwidth]{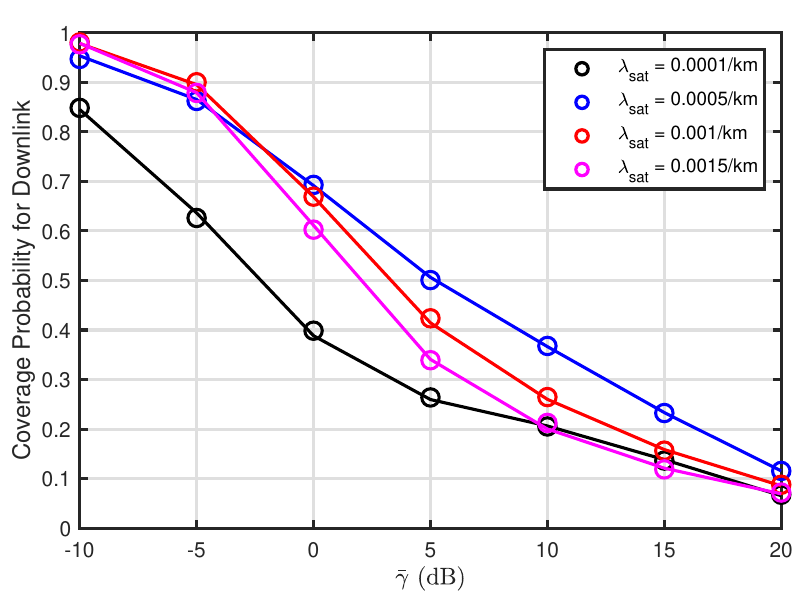}
    }
	\caption{Effects of satellite density $\lambda_{\mathrm{sat}}$ on the downlink coverage probability under STIRANs with different frequencies}
	\label{fig:lambda_sat_downlink_diff_frequencies}
\end{figure}

\begin{figure} [http]
    \centering
    \subfigure[$R_0=100\mathrm{m}$]{
	\includegraphics[width=0.427\textwidth]{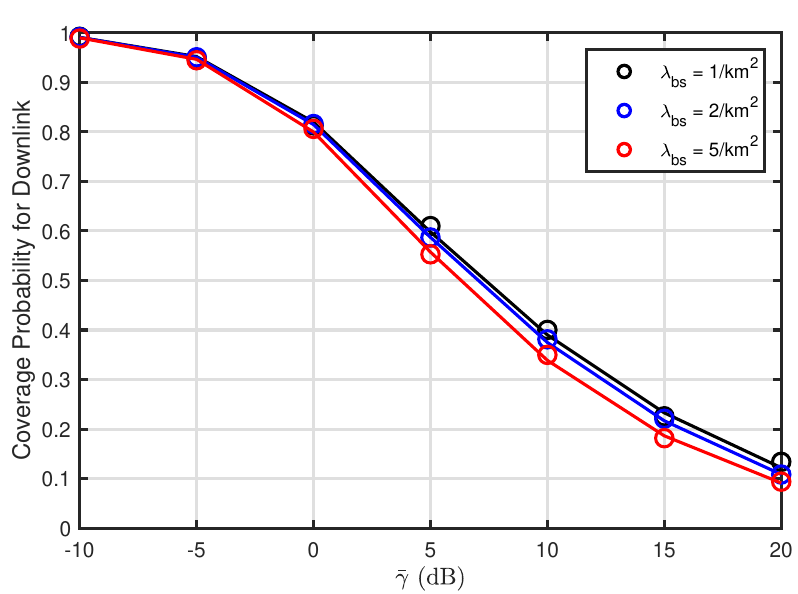}
    }
    \quad
    \subfigure[$R_0 = 700\mathrm{m}$]{
	\includegraphics[width=0.427\textwidth]{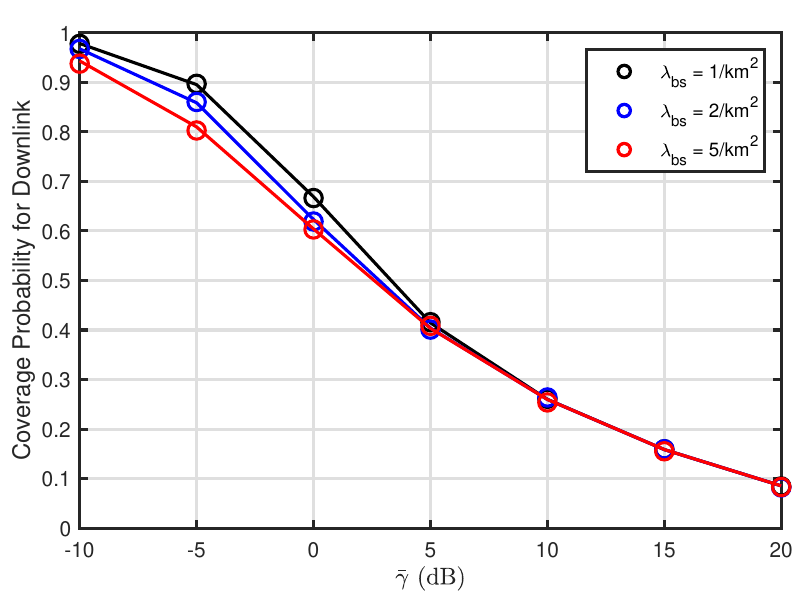}
    }
	\caption{Effects of satellite density $\lambda_{\mathrm{bs}}$ on the downlink coverage probability under STIRANs with different frequencies}
	\label{fig:lambda_bs_downlink_diff_frequencies}
\end{figure}

\begin{figure} [http]
    \centering
    \subfigure[$R_0=100\mathrm{m}$]{
	\includegraphics[width=0.427\textwidth]{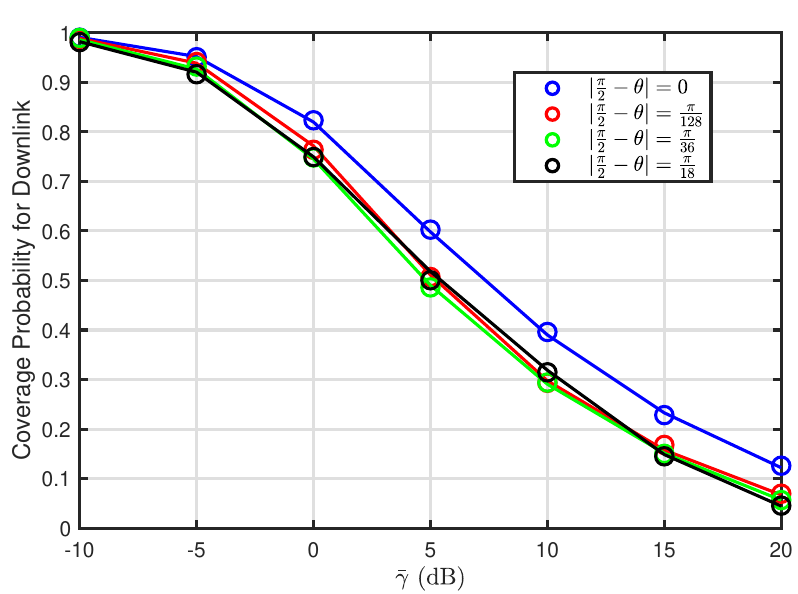}
    }
    \quad
    \subfigure[$R_0 = 700\mathrm{m}$]{
	\includegraphics[width=0.427\textwidth]{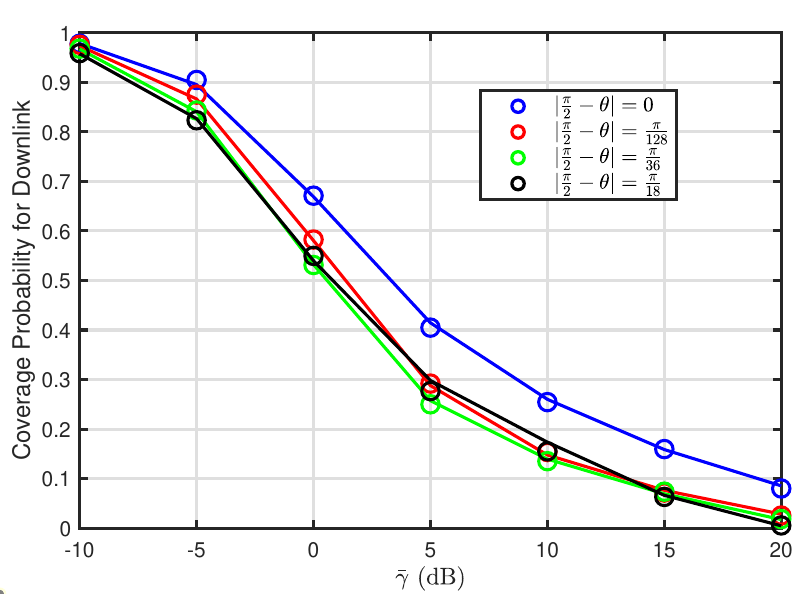}
    }
	\caption{Effects of satellite density $\theta$ on the downlink coverage probability under STIRANs with different frequencies}
	\label{fig:theta_downlink_diff_frequencies}
\end{figure}

\begin{figure} [http]
    \centering
    \subfigure[$R_0=100\mathrm{m}$]{
	\includegraphics[width=0.427\textwidth]{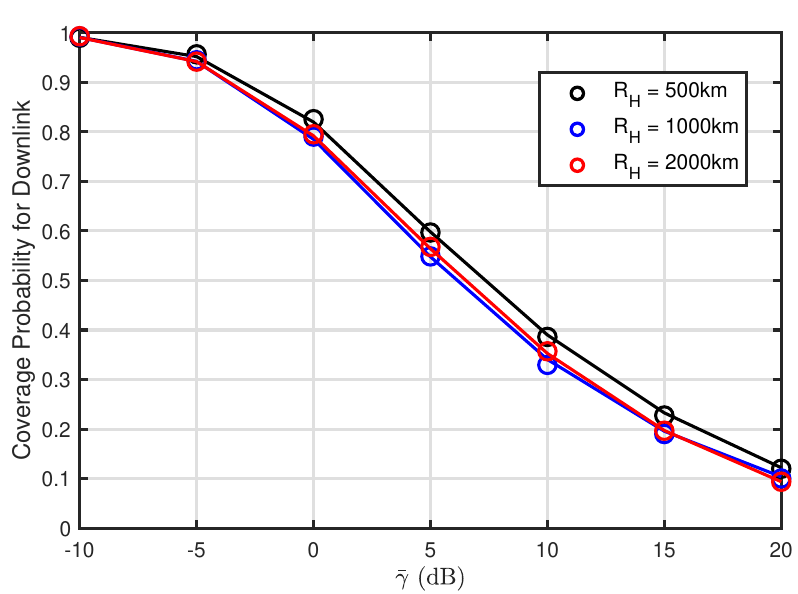}
    }
    \quad
    \subfigure[$R_0 = 700\mathrm{m}$]{
	\includegraphics[width=0.427\textwidth]{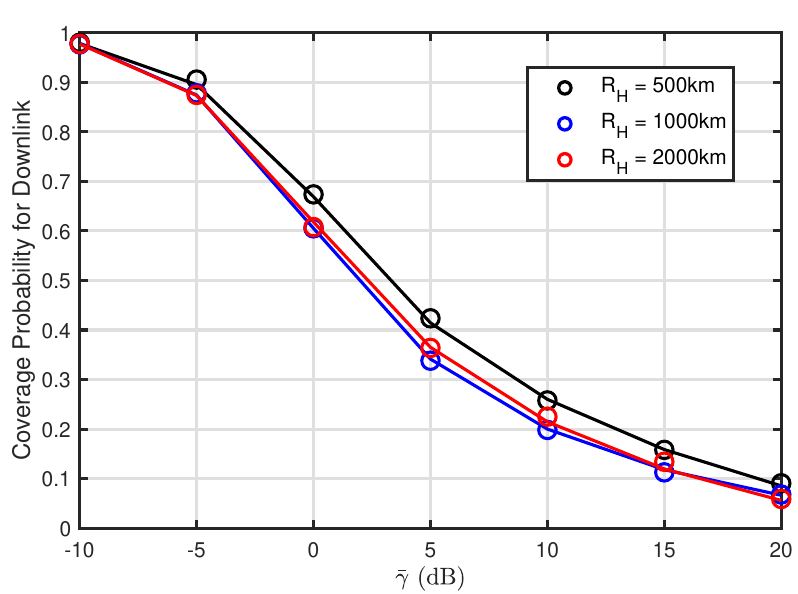}
    }
	\caption{Effects of $R_\mathrm{H}$ on the downlink coverage probability under STIRANs with different frequencies}
	\label{fig:RH_downlink_diff_frequencies}
\end{figure}

In Figs.\ref{fig:lambda_sat_downlink_diff_frequencies}, \ref{fig:lambda_bs_downlink_diff_frequencies}, \ref{fig:theta_downlink_diff_frequencies} and \ref{fig:RH_downlink_diff_frequencies}, We investigate the impact of parameters on downlink coverage probability in STIRANs with different frequencies.

In Fig.\ref{fig:lambda_sat_downlink_diff_frequencies} we analyze the effect of $\lambda_{\mathrm{sat}}$ on the downlink coverage probability under STIRANs with different frequencies. When $R_0=100\mathrm{m}$ and $R_0=700\mathrm{m}$, the coverage probability improves as $\lambda_{\mathrm{sat}}$ decreases for the high SINR region. This is because the nearest satellite distance is bounded by $r_{\textrm{min}}=\sqrt{R^2-2R_{\mathrm{E}}R \sin\left(\theta\right)+R_{\mathrm{E}}^2} $ under the proposed satellite networks, which leads to the coverage probability decreasing as the satellite density $\lambda_{\mathrm{sat}}$ increases. However, when $\lambda_{\mathrm{sat}}$ is very small, the coverage probability is extremely lower. Which is because the probability of satellite invisible becomes high, and the typical user cannot be served by satellites. In STIRANs with different frequencies scenario, $\lambda_{\mathrm{sat}}$ does not affect the terrestrial network coverage probability so STIRANs' coverage probability improves as $\lambda_{\mathrm{sat}}$ decreases.

Fig.\ref{fig:lambda_bs_downlink_diff_frequencies} shows the effect of $\lambda_{\mathrm{bs}}$ on the downlink coverage probability under STIRANs with different frequencies. As the $\lambda_{\mathrm{bs}}$ increases, the average distance from the typical user to BSs decreases, resulting in an increase in the useful signal and interference signal. Based on Fig. \ref{fig:lambda_bs_downlink_diff_frequencies}(a), when $R_0=100\mathrm{m}$, the coverage probability decreases with $\lambda_{\mathrm{bs}}$ increases in high SINR region. And in Fig. \ref{fig:lambda_bs_downlink_diff_frequencies}(b), when $R_0=700\mathrm{m}$, the coverage probability decreases with $\lambda_{\mathrm{bs}}$ increases in the low SINR region. In high SINR region, due to the distance between the typical user and the serving terrestrial BS is large, causing the useful signal becomes very weak so the coverage probability of STIRANs is primarily determined by the satellite network coverage probability.

Fig.\ref{fig:theta_downlink_diff_frequencies} illustrates the effect of $\theta$ on the downlink coverage probability. Results in Fig.\ref{fig:theta_downlink_diff_frequencies} show that when $R_0=100\mathrm{m}$ and $R_0=700\mathrm{m}$, the downlink coverage probability decreases as the $\left| \frac{\pi}{2}-\theta \right| $ increases. Due to $\left| \frac{\pi}{2}-\theta \right| $ increases, the average distance between the typical user and satellites increases. Which leads to a decrease in both the useful signal and the interference signal.
We could observe that when $\left| \theta -\frac{\pi}{2}\right| = 0$, which means the orbit passes the zenith of the user, the highest coverage performance is achievable.

In Fig. \ref{fig:RH_downlink_diff_frequencies}, we examine the effect of  $R_\mathrm{H}$ on the downlink coverage probability under STIRANs with different frequencies. We could observe that when $R_0=100\mathrm{m}$ and $R_0=700\mathrm{m}$, as the $R_\mathrm{H}$ increase the coverage probability first decreases and then increases in the high SINR region. Increasing $R_\mathrm{H}$ results the distance between the typical user and satellites increases and the number of satellites in the visible orbit increases According to the results, we could improve
the network’s coverage probability by optimizing $R_\mathrm{H}$.
\subsection{The downlink probability of the STIRANs with the same frequency}
\subsubsection{Association Probability}
\begin{figure} [http]
    \centering
    \subfigure[The influence of $\lambda_{\mathrm{sat}}$ and $\lambda_{\mathrm{bs}}$ on the association probability]{ \includegraphics[width=0.427\textwidth]{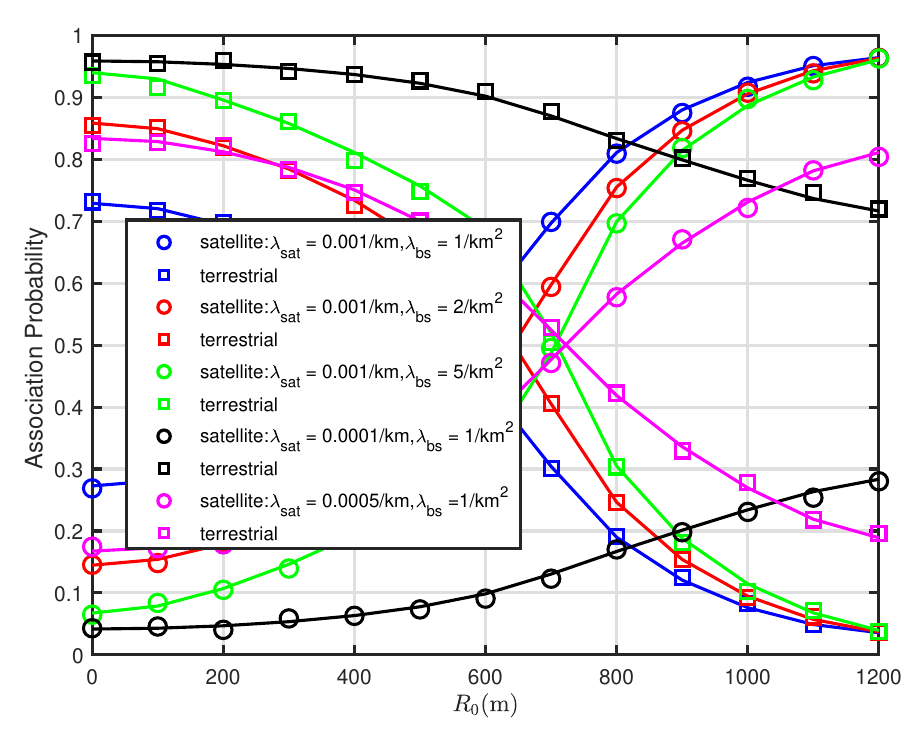}
    }
    \quad
    \subfigure[The influence of $\theta$ on the association probability]{
	\includegraphics[width=0.427\textwidth]{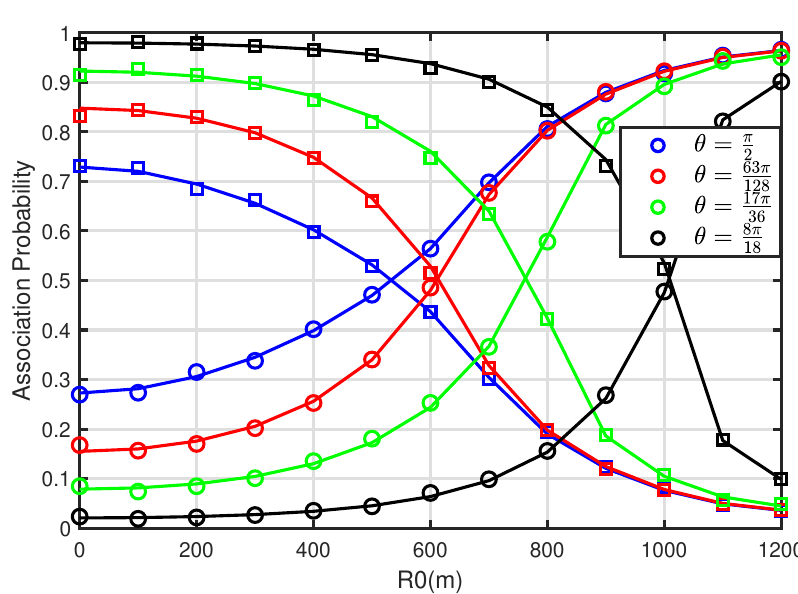}
    }
	\caption{The association probability under different $R_0$}
	\label{fig:Association_probability}
\end{figure}

In Fig.\ref{fig:Association_probability} illustrates the impact of network parameters on the association probability. The derived expressions of the association probability perfectly match the simulation results. As we expect, with an increase in $R_0$, the association probability with the terrestrial BS decreases, while the association probability with the satellite increases. This is because the received power from the serving terrestrial BS decreases as the distance $R_0$ increases and eventually tends to be flat.

In Fig.\ref{fig:Association_probability}(a) shows the relationship between the association probability and $R_0$ under different $\lambda_{\mathrm{sat}}$ and $\lambda_{\mathrm{bs}}$. An increase in $\lambda_{\mathrm{sat}}$ leads to a higher probability of associating with satellites and a lower probability of associating with terrestrial BSs. Conversely, an increase in $\lambda_{\mathrm{bs}}$ results in a higher probability of associating with terrestrial BSs and a lower probability of associating with satellites. This is because as BS density increases, the average distance from the typical user to the serving BS decreases, and the useful signal power received by the typical user is stronger, leading to an increased association probability.

And in Fig.\ref{fig:Association_probability}(b), we depict the relationship between the association probability and $\theta$. With the $\theta$ decrease, the probability of associating with the terrestrial BS increases, and the probability of associating with the satellite decreases. This is because when $\theta$ decreases, the distance from the typical users to the serving satellite will also increase, resulting in the received power from the serving satellite being smaller. Therefore, the typical user prefers to associate with terrestrial BS.
\subsubsection{Downlink Probability}

\begin{figure} [http]
    \centering
    \subfigure[$R_0=100\mathrm{m}$]{
	\includegraphics[width=0.427\textwidth]{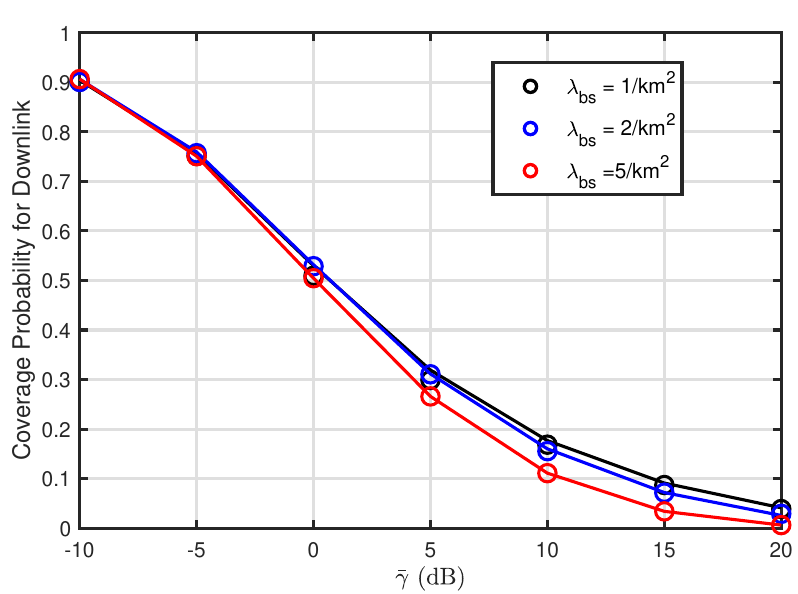}
    }
    \quad
    \subfigure[$R_0 = 700\mathrm{m}$]{
	\includegraphics[width=0.427\textwidth]{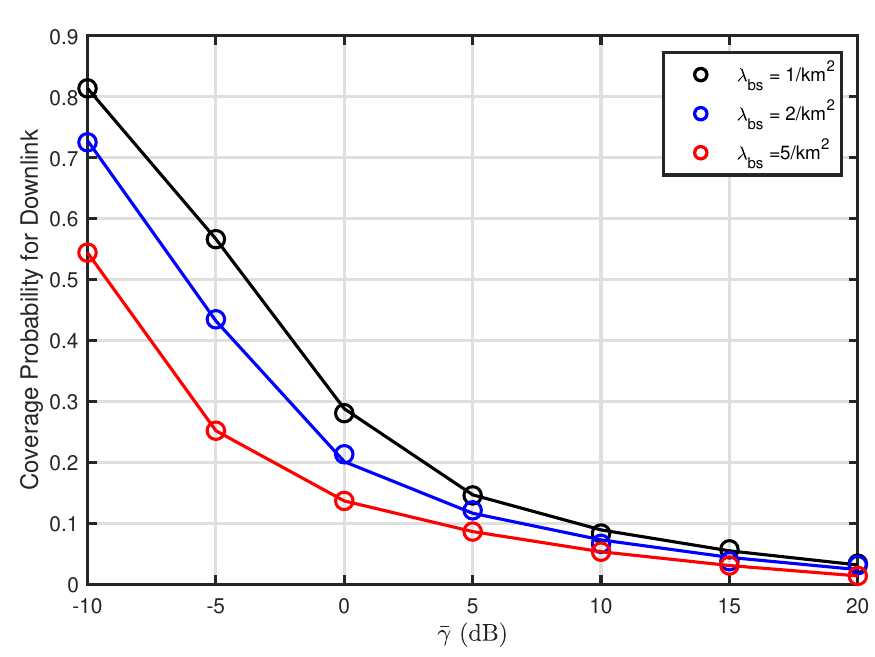}
    }
	\caption{Effects of $\lambda_{\mathrm{bs}}$ on the downlink coverage probability under STIRANs with same frequency}
	\label{fig:lambda_bs_downlink_same_frequency}
\end{figure}

\begin{figure} [http]
	\centering
    \subfigure[ $R_0 = 100\mathrm{m}$]{
	\includegraphics[width=0.427\textwidth]{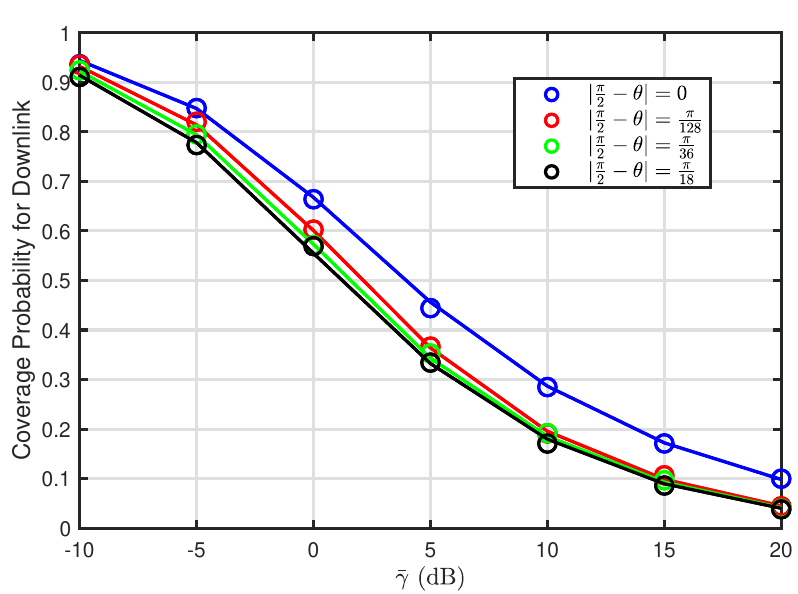}
    }
    \quad
    \subfigure[$R_0 = 700\mathrm{m}$ ]{
	\includegraphics[width=0.427\textwidth]{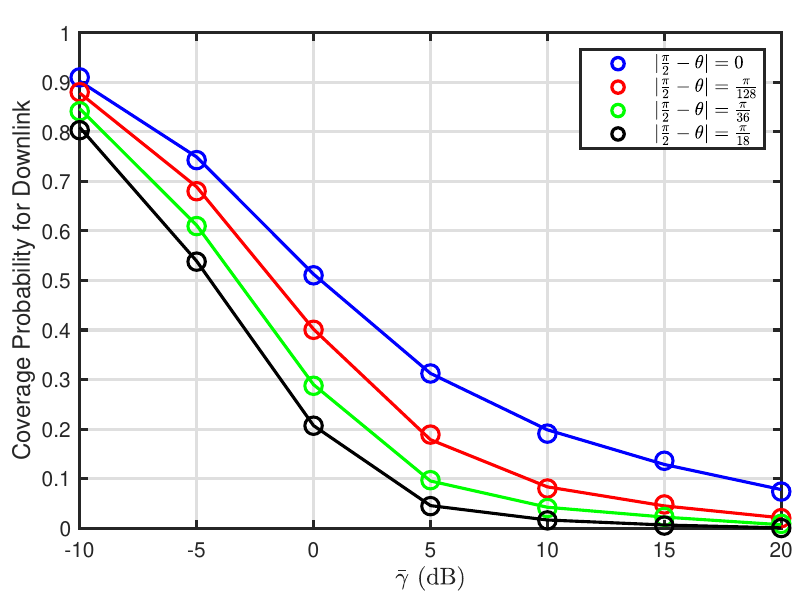}
    }
	\caption{Effects of satellite polar angle $\theta$ on the downlink coverage probability under STIRANs with same frequency}
	\label{fig:theta_downlink_same}
\end{figure}

\begin{figure} [http]
	\centering
    \subfigure[ $R_0 = 100\mathrm{m}$]{
	\includegraphics[width=0.427\textwidth]{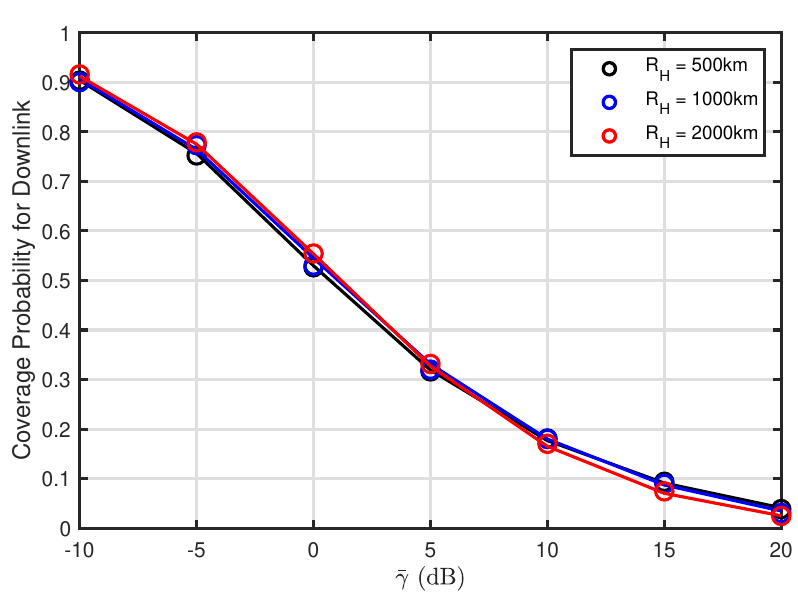}
    }
    \quad
    \subfigure[$R_0 = 700\mathrm{m}$ ]{
	\includegraphics[width=0.427\textwidth]{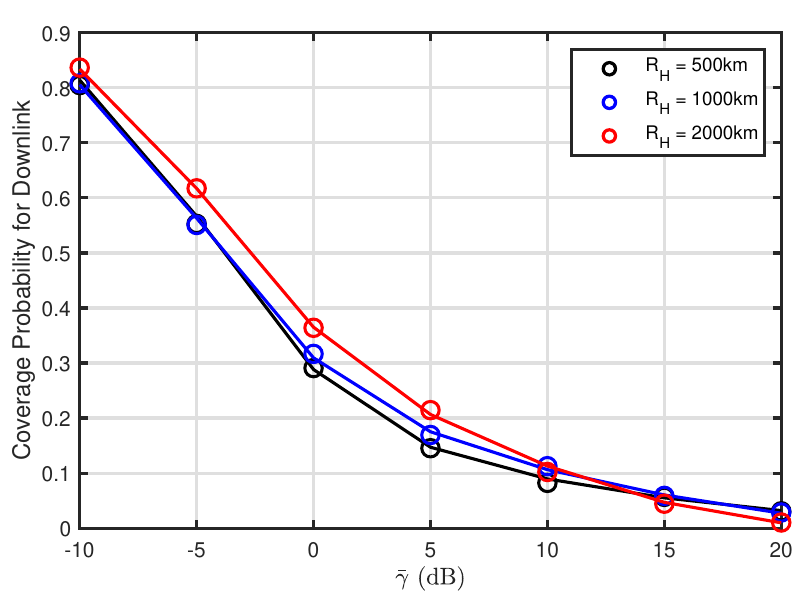}
    }
	\caption{Effects of $R_\mathrm{H}$ on the downlink coverage probability under STIRANs with same frequency}
	\label{fig:RH_downlink_same}
\end{figure}

In Figs.\ref{fig:lambda_sat_downlink_same_frequency}, \ref{fig:lambda_bs_downlink_same_frequency}, \ref{fig:theta_downlink_same} and \ref{fig:RH_downlink_same}, we investigate the impact of parameters on downlink coverage probability in STIRANs under a setting of the same frequency. The results show that the accuracy of our analysis is verified through extensive simulation evaluation.
\begin{figure} [http]\label{fig:lambda_sat_downlink_same_frequency}
    \centering
    \subfigure[$R_0=100\mathrm{m}$]{
	\includegraphics[width=0.427\textwidth]{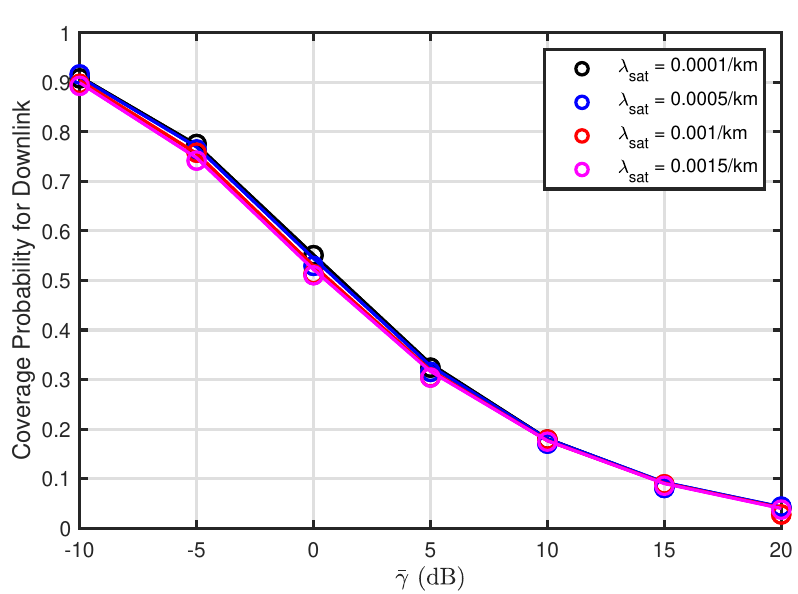}
    }
    \quad
    \subfigure[$R_0 = 700\mathrm{m}$]{
	\includegraphics[width=0.427\textwidth]{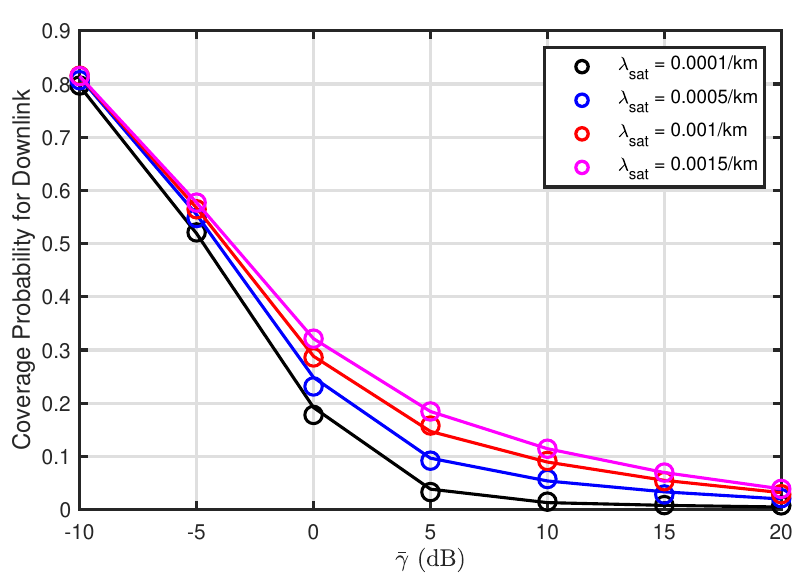}
    }
	\caption{Effects of satellite density $\lambda_{\mathrm{sat}}$ on the downlink coverage probability under STIRANs with same frequency}
	\label{fig:lambda_sat_downlink_same_frequency}
\end{figure}

Fig.\ref{fig:lambda_sat_downlink_same_frequency} analyzes the effect of $\lambda_{\mathrm{sat}}$ on the downlink coverage probability in STIRANs under a setting of same frequency. In Fig.\ref{fig:lambda_sat_downlink_same_frequency}(a), we could know that when $R_0=100\mathrm{m}$, the coverage probability does not change significantly as $\lambda_{\mathrm{sat}}$ increases. In Fig.\ref{fig:lambda_sat_downlink_same_frequency}(b), $R_0=700\mathrm{m}$, as $\lambda_{\mathrm{sat}}$ increases, the coverage probability also increases. Which is because as the $\lambda_{\mathrm{sat}}$ increases, the average distance between typical users and satellites decreases, leading to a decrease in path loss. This involves both the received power from the serving satellite and interference from other satellites. When $R_0=700\mathrm{m}$ and $\lambda_{\mathrm{sat}}$ is low, the impact of the received power from the serving satellite is more significant.

Fig.\ref{fig:lambda_bs_downlink_same_frequency} analyzes the effect of $\lambda_{\mathrm{bs}}$ on the downlink coverage probability in STIRANs with the same frequency. In Fig.\ref{fig:lambda_bs_downlink_same_frequency}(a), $R_0=100\mathrm{m}$, the downlink coverage probability decreases as $\lambda_{\mathrm{bs}}$ rises in the high SINR region. This is because the average distance between the typical user and the terrestrial BSs decreases. As a result, both the desired serving signal and the interfering signal larger and interference from the terrestrial BSs becomes more influential. This leads to a faster decline rate of the coverage probability as $\bar{\gamma}$ increases. When $R_0=700\mathrm{m}$, as $\lambda_{\mathrm{bs}}$ increases,the coverage probability decreases. The coverage probability decreases due to the typical user prefers to associate with the serving satellite, so terrestrial BSs as the interference increases with the $\lambda_{\mathrm{bs}}$ increases.

Fig.\ref{fig:theta_downlink_same} examine how the orbit polar angle $\theta$ effects the donwlink coverage probability under STIRANs with same frequency. Results in Fig.\ref{fig:theta_downlink_same} show that when $R_0=100\mathrm{m}$ and $R_0=700\mathrm{m}$, the donwlink coverage probability decreases as the $\left| \frac{\pi}{2}-\theta \right| $ increases. As we can seen in this figure, when $\left| \frac{\pi}{2}-\theta \right| =0 $, the orbit passes the zenith of the user, the highest coverage performance is achievable.

Fig.\ref{fig:RH_downlink_same} shows the effect of $R_\mathrm{H}$ on the downlink coverage probability under STIRANs with same frequencies.
 we could observe that when $R_0=100\mathrm{m}$, the coverage probability does not change significantly as the $R_\mathrm{H}$ increases.
 When $R_0=700\mathrm{m}$, the coverage probability  increases with the increase in $R_H$.

\section{Conclusion} \label{conclusion}
In this paper, we have studied the downlink probability in STIRANs under the settings of same frequency and distinct frequencies. To analyze the impact of satellite orbit geometrical parameters on performance, we have modeled satellite positions as a one-dimensional PPP and terrestrial BSs as a homogeneous PPP, taking into account coverage holes in the terrestrial scenario. Then, we have derived downlink coverage probability expressions in terms of network design parameters under STIRANs with different frequencies and the same frequency. The accuracy of the derivations has been confirmed by numerical results. Simulation results have shown that STIRANs can effectively resolve coverage holes in terrestrial networks.


\begin{thebibliography}{99}
\bibitem{34}X. Zhu and C. Jiang, "Integrated Satellite-Terrestrial Networks Toward 6G: Architectures, Applications, and Challenges," in IEEE Internet of Things Journal, vol. 9, no. 1, pp. 437-461, 1 Jan.1, 2022.
\bibitem{33}Q. Chen, W. Meng, S. Han, C. Li and T. Q. S. Quek, "Coverage Analysis of SAGIN With Sectorized Beam Pattern Under Shadowed-Rician Fading Channels," in IEEE Transactions on Communications, vol. 71, no. 8, pp. 4988-5004, Aug. 2023.
\bibitem{40}R. Wang, M. A. Kishk and M. -S. Alouini, "Stochastic Geometry-Based Low Latency Routing in Massive LEO Satellite Networks," in IEEE Transactions on Aerospace and Electronic Systems, vol. 58, no. 5, pp. 3881-3894, Oct. 2022
\bibitem{31}Y. Sun, M. Peng, S. Zhang, G. Lin and P. Zhang,"Integrated Satellite-Terrestrial Networks: Architectures, Key Techniques, and Experimental Progress," in IEEE Network, vol. 36, no. 6, pp. 191-198.
\bibitem{35}I. Portillo, B. G. Cameron and E. F. Crawley, “A technical comparison of three low earth orbit satellite constellation systems to provide global broadband,” Acta Astronautica, 2019, 159: 123-135.
\bibitem{36}J. Sedin, L. Feltrin and X. Lin, "Throughput and capacity evaluation of 5G new radio non-terrestrial networks with LEO satellites," in Proc. 2020 IEEE GLOBECOM, Taipei, 2020: 1-6.
\bibitem{1}J. G. Andrews, F. Baccelli and R. K. Ganti, "A Tractable Approach to Coverage and Rate in Cellular Networks," in IEEE Transactions on Communications, vol. 59, no. 11, pp. 3122-3134, November 2011
\bibitem{3}M. Ding, P. Wang, D. López-Pérez, G. Mao and Z. Lin, "Performance Impact of LoS and NLoS Transmissions in Dense Cellular Networks," in IEEE Transactions on Wireless Communications, vol. 15, no. 3, pp. 2365-2380, March 2016
\bibitem{4}H. S. Dhillon, R. K. Ganti, F. Baccelli and J. G. Andrews, "Modeling and Analysis of K-Tier Downlink Heterogeneous Cellular Networks," in IEEE Journal on Selected Areas in Communications, vol. 30, no. 3, pp. 550-560, April 2012
\bibitem{5}M. D. Renzo, A. Guidotti and G. E. Corazza, "Average Rate of Downlink Heterogeneous Cellular Networks over Generalized Fading Channels: A Stochastic Geometry Approach," in IEEE Transactions on Communications, vol. 61, no. 7, pp. 3050-3071, July 2013
\bibitem{2}T. D. Novlan, H. S. Dhillon and J. G. Andrews, "Analytical Modeling of Uplink Cellular Networks," in IEEE Transactions on Wireless Communications, vol. 12, no. 6, pp. 2669-2679, June 2013
\bibitem{6}H. ElSawy and E. Hossain, "On Stochastic Geometry Modeling of Cellular Uplink Transmission With Truncated Channel Inversion Power Control," in IEEE Transactions on Wireless Communications, vol. 13, no. 8, pp. 4454-4469, Aug. 2014
\bibitem{7}]P. Herath, C. Tellambura and W. A. Krzymien, "Stochastic Geometry Modeling of Cellular Uplink Power Control under Composite Rayleigh-Lognormal Fading," 2015 IEEE 82nd Vehicular Technology Conference (VTC2015-Fall), Boston, MA, USA, 2015, pp. 1-5,
\bibitem{8}N. Okati, T. Riihonen, D. Korpi, I. Angervuori and R. Wichman, "Downlink Coverage and Rate Analysis of Low Earth Orbit Satellite Constellations Using Stochastic Geometry," in IEEE Transactions on Communications, vol. 68, no. 8, pp. 5120-5134, Aug. 2020
\bibitem{9}N. Okati and T. Riihonen, "Stochastic Analysis of Satellite Broadband by Mega-Constellations with Inclined LEOs," 2020 IEEE 31st Annual International Symposium on Personal, Indoor and Mobile Radio Communications, London, UK, 2020, pp. 1-6
\bibitem{10}A. Talgat, M. A. Kishk and M. -S. Alouini, "Nearest Neighbor and Contact Distance Distribution for Binomial Point Process on Spherical Surfaces," in IEEE Communications Letters, vol. 24, no. 12, pp. 2659-2663, Dec. 2020
\bibitem{11}LEO Satellite Communication Systems," in IEEE Communications Letters, vol. 25, no. 8, pp. 2458-2462, Aug. 2021
\bibitem{12}D. -H. Jung, J. -G. Ryu, W. -J. Byun and J. Choi, " Performance Analysis of Satellite Communication System Under the Shadowed-Rician Fading: A Stochastic Geometry Approach," in IEEE Transactions on Communications, vol. 70, no. 4, pp. 2707-2721, April 2022
\bibitem{13}R. Wang, P. Ren, D. Xu and L. Lu, "Stochastic Geometry Analysis of LEO Constellation Coverage under Atmospheric Attenuation," 2022 IEEE 96th Vehicular Technology Conference (VTC2022-Fall), London, United Kingdom, 2022, pp. 1-5
\bibitem{14}R. Wang, M. A. Kishk and M. -S. Alouini, "Evaluating the Accuracy of Stochastic Geometry Based Models for LEO Satellite Networks Analysis," in IEEE Communications Letters, vol. 26, no. 10, pp. 2440-2444, Oct. 2022
\bibitem{15}A. Al-Hourani, “An analytic approach for modeling the coverage performance of dense satellite networks,” IEEE Wireless Commun. Lett., vol. 10, no. 4, pp. 897–901, Apr. 2021.
\bibitem{16}G. Kim, S. Lee, H. Lim, B. C. Jung and S. H. Chae, "Coverage Probability Analysis of LEO Satellite Communication Systems With Directional Beamforming," 2023 Fourteenth International Conference on Ubiquitous and Future Networks (ICUFN), Paris, France, 2023, pp. 243-247
\bibitem{17}S. H. Chae, H. Lim, H. Lee and B. C. Jung, "Performance analysis of dense low earth orbit satellite communication networks with stochastic geometry," in Journal of Communications and Networks, vol. 25, no. 2, pp. 208-221, April 2023
\bibitem{18}J. Park, J. Choi and N. Lee, "A Tractable Approach to Coverage Analysis in Downlink Satellite Networks," in IEEE Transactions on Wireless Communications, vol. 22, no. 2, pp. 793-807, Feb. 2023
\bibitem{19}N. Lee, "Coverage Analysis for Satellite Downlink Networks," ICC  2022 - IEEE International Conference on Communications, Seoul, Korea, Republic of, 2022, pp. 2387-2392
\bibitem{20}Y. Sun and Z. Ding, "A Fine Grained Stochastic Geometry Based Analysis on LEO Satellite Communication Systems," in IEEE Networking Letters
\bibitem{21}N. Okati and T. Riihonen, "Modeling and Analysis of LEO Mega-Constellations as Nonhomogeneous Poisson Point Processes," 2021 IEEE 93rd Vehicular Technology Conference (VTC2021-Spring), Helsinki, Finland, 2021, pp. 1-5
\bibitem{22}N. Okati and T. Riihonen, "Nonhomogeneous stochastic geometry analysis of massive LEO communication constellations", IEEE Transactions on Communications, Jan. 2022.
\bibitem{23}A. Yastrebova et al., "Theoretical and Simulation-based Analysis of Terrestrial Interference to LEO Satellite Uplinks," GLOBECOM 2020 - 2020 IEEE Global Communications Conference, Taipei, Taiwan, 2020, pp. 1-6
\bibitem{24}H. Jia, C. Jiang, L. Kuang and J. Lu, "An Analytic Approach for Modeling Uplink Performance of Mega Constellations," in IEEE Transactions on Vehicular Technology, vol. 72, no. 2, pp. 2258-2268, Feb. 2023
\bibitem{25}Lee, J. Noh, S.`                                                                                                                                                                                                                                                                                                                                                                                                                                                                                                                                                                                                                                                                                                                                                                                                                                                                                                                                                                                                                                                                                                                                                                                                                                                                                                                                                                                                                                                                                                                                                                                                                                                                                                                                                                                                                                                                                                                                                                                                                                                                                                                                                                                                                                                                                                                                                                                                                                                                                                                                                                                                                                                                                                                                                                                                                                                                                                                                                                                                                                                                                                                                                                                                                                                                                                                                                                                                                                                                                                                                                                                                                                                                                                                                                                                                                                                                                                                                                                                                                                                                                                                                                                                                                                                                                                                                                                                                                                                                                                                                                                                                                                                                                                                                                                                                                                                                                                                                                                                                                                                                                                                                                                                                                                                                                                                                                                                                                                                                                                                                                                                                                                                                                                                        ` Jeong, S. Lee, Namyoon. (2022). Coverage Analysis of LEO Satellite Downlink Networks: Orbit Geometry Dependent Approach. 10.48550/arXiv.2206.09382.
\bibitem{26}Y. Tian, G. Pan, M. A. Kishk and M. -S. Alouini, "Stochastic Analysis of Cooperative Satellite-UAV Communications," in IEEE Transactions on Wireless Communications, vol. 21, no. 6, pp. 3570-3586, June 2022.
\bibitem{27}J. Kokkoniemi, J. M. Jornet, V. Petrov, Y. Koucheryavy, and M. Juntti,“Channel modeling and performance analysis of airplane-satellite terahertz band communications,” IEEE Trans. Veh. Technol., vol. 70,no. 3, pp. 2047–2061, Mar. 2021.
\bibitem{28}X. Wang, N. Deng and H. Wei, "Coverage and Rate Analysis of LEO Satellite-to-Airplane Communication Networks in Terahertz Band," in IEEE Transactions on Wireless Communications, vol. 22, no. 12, pp. 9076-9090, Dec. 2023.
\bibitem{29}O. Y. Kolawole, S. Vuppala, "On the Performance of Cognitive Satellite-Terrestrial Networks," in IEEE Transactions on Cognitive Communications and Networking, vol. 3, no. 4, pp. 668-683, Dec. 2017.
\bibitem{30}B. A. Homssi and A. Al-Hourani, "Modeling Uplink Coverage Performance in Hybrid Satellite-Terrestrial Networks," in IEEE Communications Letters, vol. 25, no. 10, pp. 3239-3243, Oct. 2021
\bibitem{37}F. Baccelli and B. Blaszczyszyn, “Stochastic geometry and wireless networks: Volume i theory,” Found. Trends in Networking, vol. 3, no. 3–4, pp. 249–449, Mar. 2009.
\bibitem{38}A. Abdi, W. Lau, M.-S. Alouini, and M. Kaveh, “A new simple model for land mobile satellite channels: first- and second-order statistics,” IEEE Trans. Wireless Commun., vol. 2, no. 3, pp. 519–528, 2003.
\bibitem{39}Satellite antenna radiation patterns for non-geostationary orbit satellite antennas operating in the fixed-satellite service below 30GHz, Recommendation ITU-R S.1528, 2001
\end{thebibliography}
\end{document}